\documentclass[aps,nofootinbib,prx,superscriptaddress,tightenlines,notitlepage,twocolumn,showpacs,longbibliography]{revtex4-1}

\usepackage[colorlinks]{hyperref}
\usepackage{tabularx}
\usepackage{graphicx}
\usepackage{amssymb,bm,tensor}
\usepackage{xcolor}
\usepackage{amsmath}
\usepackage[varg]{txfonts}
\usepackage{enumerate}
\usepackage[normalem]{ulem}

\newcommand{\AEI}{\affiliation{Max Planck Institute for Gravitational Physics
(Albert Einstein Institute), Am M\"uhlenberg 1, Potsdam D-14476, Germany}}
\newcommand{\MPIfR}{\affiliation{Max-Planck-Institut f\"ur Radioastronomie, Auf
dem H\"ugel 69, D-53121 Bonn, Germany}}
\newcommand{\Maryland}{\affiliation{Department of Physics, University of
Maryland, College Park, MD 20742, USA}}
\newcommand{\Manchester}{\affiliation{Jodrell Bank Centre for Astrophysics, The
University of Manchester, M13 9PL, United Kingdom}}
\newcommand{\enDash}{\mbox{--}}

\begin{document}

\title{Constraining nonperturbative strong-field effects in scalar-tensor
gravity by combining pulsar timing and laser-interferometer gravitational-wave
detectors}

\author{Lijing Shao}\thanks{lijing.shao@aei.mpg.de}
\AEI

\author{Noah Sennett}
\AEI
\Maryland

\author{Alessandra Buonanno}
\AEI
\Maryland

\author{Michael Kramer}
\MPIfR
\Manchester

\author{Norbert Wex}
\MPIfR

\date{\today}

\begin{abstract}
  Pulsar timing and laser-interferometer gravitational-wave (GW) detectors are
  superb laboratories to study gravity theories in the strong-field regime.
  Here we combine those tools to test the mono-scalar-tensor theory of Damour
  and Esposito-Far{\`e}se (DEF), which predicts nonperturbative scalarization
  phenomena for neutron stars (NSs).    First, applying Markov-chain Monte
  Carlo techniques, we use the absence of dipolar radiation in the
  pulsar-timing observations of five binary systems composed of a NS and a
  white dwarf, and eleven equations of state (EOSs) for NSs, to derive the most
  stringent constraints on the two free parameters of the DEF scalar-tensor
  theory. Since the binary-pulsar bounds depend on the NS mass and the EOS, we
  find that current pulsar-timing observations leave {\it scalarization
  windows}, i.e., regions of parameter space where scalarization can still be
  prominent.  Then, we investigate if these scalarization windows could be
  closed and if pulsar-timing constraints could be improved by
  laser-interferometer GW detectors, when spontaneous (or dynamical)
  scalarization sets in during the early (or late) stages of a binary NS (BNS)
  evolution. For the early inspiral of a BNS carrying constant scalar charge,
  we employ a Fisher matrix analysis to show that Advanced LIGO can improve
  pulsar-timing constraints for some EOSs, and  next-generation detectors, such
  as the Cosmic Explorer and Einstein Telescope, will be able to improve those
  bounds for all eleven EOSs.  Using the late inspiral of a BNS, we estimate
  that for some of the EOSs under consideration the onset of dynamical
  scalarization can happen early enough to improve the constraints on the DEF
  parameters obtained by combining the five binary pulsars. Thus, in the near
  future the complementarity of pulsar timing and direct observations of GWs on
  the ground  will be extremely valuable in probing gravity theories in the
  strong-field regime.
\end{abstract}

\maketitle

%% === main body of the paper ===

%-------------------------------------------------------------------------------
\section{Introduction}
\label{sec:intro}
%-------------------------------------------------------------------------------

In general relativity (GR), gravity is mediated solely by a rank-2 tensor,
namely the spacetime metric $g_{\mu\nu}$. Scalar-tensor theories of gravity,
which add a scalar component to the gravitational interaction, are popular
alternatives to GR. Though first proposed in 1921~\cite{Kaluza:1921tu},
contemporary interest in these theories has been spurred by their potential
connection to inflation and dark energy, as well as possible unified theories
of quantum gravity~\cite{Clifton:2011jh}.  A modern framework for the class of
scalar-tensor theories we consider was developed in Refs.~\cite{Jordan:1949zz,
Jordan:1952schwerkraft, Fierz:1956zz, Brans:1961sx, Damour:1992we, Will:1993ns}
(see also more generic Horndeski scalar-tensor theories in
Ref.~\cite{Horndeski:1974wa}).

Ultimately, the existence (or absence) of scalar degrees of freedom in gravity
will be decided by experiments. Most scalar-tensor theories are designed to be
{\it metric theories of gravity}, that is, they respect the Einstein
equivalence principle~\cite{Will:1993ns, Will:2014kxa}.  Therefore, precision
tests of the weak-equivalence principle, the local Lorentz invariance, and the
local position invariance in flat spacetime are unable to constrain
them~\cite{Will:1993ns, Will:2014kxa, Berti:2015itd, Shao:2016ezh}. However,
such theories generally violate the strong-equivalence principle. Tests of the
strong-equivalence principle with self-gravitating bodies provide an ideal
window to experimentally search for (or rule out) the scalar sector of
gravity~\cite{Will:1993ns, Will:2014kxa, Shao:2016ubu}.

Particularly prominent violations of the strong-equivalence principle are known
to arise in the class of massless mono-scalar-tensor theories, studied by
Damour and Esposito-Far{\` e}se in the form of {\it nonperturbative
strong-field effects} in neutron stars (NSs)~\cite{Damour:1993hw,
Damour:1996ke, Damour:1998jk}.  In this paper, we investigate the extent to
which pulsar timing and ground-based gravitational-wave (GW) observations can
constrain these phenomena (space-based GW experiments~\cite{Berti:2004bd,
Berti:2005qd, Yagi:2009zz} are beyond the scope of this paper). Our results
demonstrate that, depending on the parameters of binary systems and NS
equations of state (EOSs), these two types of experiments can provide
complementary bounds on scalar-tensor theories~\cite{Will:2014kxa,
Shibata:2013pra, Berti:2015itd, Yunes:2016jcc, Shao:2016ezh}.  These results
are especially timely as new instruments come online in the upcoming years in
both fields~\cite{Shao:2014wja, Evans:2016mbw}.

The paper is organized as follows. In the next section, we briefly review two
nonperturbative phenomena, notably spontaneous
scalarization~\cite{Damour:1993hw, Damour:1996ke} and dynamical
scalarization~\cite{Barausse:2012da, Palenzuela:2013hsa, Shibata:2013pra,
Taniguchi:2014fqa, Sennett:2016rwa}, that arise in certain scalar-tensor
theories of gravity. Then, in Sec.~\ref{sec:bp}, we derive stringent
constraints on these theories by combining state-of-the-art pulsar observations
of five NS-white dwarf (WD) systems.  In Sec.~\ref{sec:gw}, we employ these
constraints and investigate the potential detectability of nonperturbative
effects in binary NS (BNS) systems using the Advanced Laser Interferometer
Gravitational-wave Observatory (LIGO)~\cite{TheLIGOScientific:2014jea} and
next-generation ground-based detectors.  Finally, in Sec.~\ref{sec:con}, we
discuss the main results and implications of our finding, and give perspectives
for future observations.

%-------------------------------------------------------------------------------
\section{Nonperturbative strong-field phenomena in scalar-tensor gravity}
\label{sec:stg}
%-------------------------------------------------------------------------------

In this paper we focus on the class of mono-scalar-tensor theories that are
defined by the following  action in the
Einstein-frame~\cite{Jordan:1952schwerkraft, Fierz:1956zz, Damour:1993hw,
Damour:1996ke},
\begin{eqnarray}
  S &=& \frac{c^4}{16 \pi G_*} \int \frac{{\rm d}^4 x}{c} \sqrt{-g_*}
  \left[ R_* - 2 g^{\mu\nu}_* \partial_\mu \varphi \partial_\nu \varphi
  - V(\varphi) \right] \nonumber\\
  && \qquad + S_m \left[ \psi_m; A^2(\varphi) g^*_{\mu\nu} \right] \,,
  \label{eq:action}
\end{eqnarray}
where $G_*$ is the bare gravitational coupling constant, $g^*_{\mu\nu}$ is the
Einstein metric with its determinant $g_*$, ${R_* \equiv g_*^{\mu\nu}
R_{\mu\nu}^*}$ is the Ricci scalar, $\psi_m$ collectively denotes the matter
content, and $A(\varphi)$ is the (conformal) coupling function that depends on
the scalar field, $\varphi$. Henceforth, for simplicity, we assume that the
potential, $V(\varphi)$, is a slowly varying function that changes on scales
much larger than typical length scales of the system that we consider, thus, we
set $V(\varphi) = 0$ in our calculation.

The field equations are derived with the least-action
principle~\cite{Damour:1992we, Will:1993ns} for $g^*_{\mu\nu}$ and $\varphi$,
\begin{eqnarray}
  R^*_{\mu\nu} &=& 2 \partial_\mu \varphi \partial_\nu \varphi
  + \frac{8\pi G_*}{c^4} \left(T^*_{\mu\nu}
  - \frac{1}{2} T^* g^*_{\mu\nu} \right) \,,
  \label{eq:field:metric} \\
  \square_{g^*} \varphi &=& -\frac{4\pi G_*}{c^4} \alpha(\varphi) T_* \,,
  \label{eq:field:scalar}
\end{eqnarray}
with the energy-momentum tensor of matter fields, $T_*^{\mu\nu} \equiv 2c
\left(-g_*\right)^{-1/2}  \delta S_m / \delta g^*_{\mu\nu} $, and the
field-dependent coupling strength between the scalar field and the trace of the
energy-momentum tensor of matter fields, $\alpha(\varphi) \equiv \partial \ln
A(\varphi) / \partial \varphi$.

Following Damour and Esposito-Far\`ese~\cite{Damour:1992we, Damour:1996ke}, we
consider a polynomial form for $\ln A(\varphi)$ up to quadratic order, that is
$A(\varphi)= \exp\left(\beta_0 \varphi^2 / 2 \right)$, and denote $\alpha_0
\equiv \alpha(\varphi_0) = \beta_0 \varphi_0$ with $\varphi_0$ the asymptotic
value of $\varphi$ at infinity. This particular scalar-tensor theory
(henceforth, DEF theory) is completely characterized by two parameters
$(\alpha_0, \beta_0)$ and for systems dominated by strong-field gravity, such
as NSs, can give rise to potentially observable, nonperturbative physical
phenomena~\cite{Damour:1993hw, Barausse:2012da}. Weak-field Solar-system
experiments, generally, only probe the $\alpha_0$-dimension or the combination
$\beta_0\alpha_0^2$ in the $(\alpha_0, \beta_0)$ parameter space~(see
Refs.~\cite{Damour:2007uf, Will:2014kxa} and references therein). In GR,
$\alpha_0 = \beta_0 = 0$.

\begin{figure}
  \includegraphics[width=9cm]{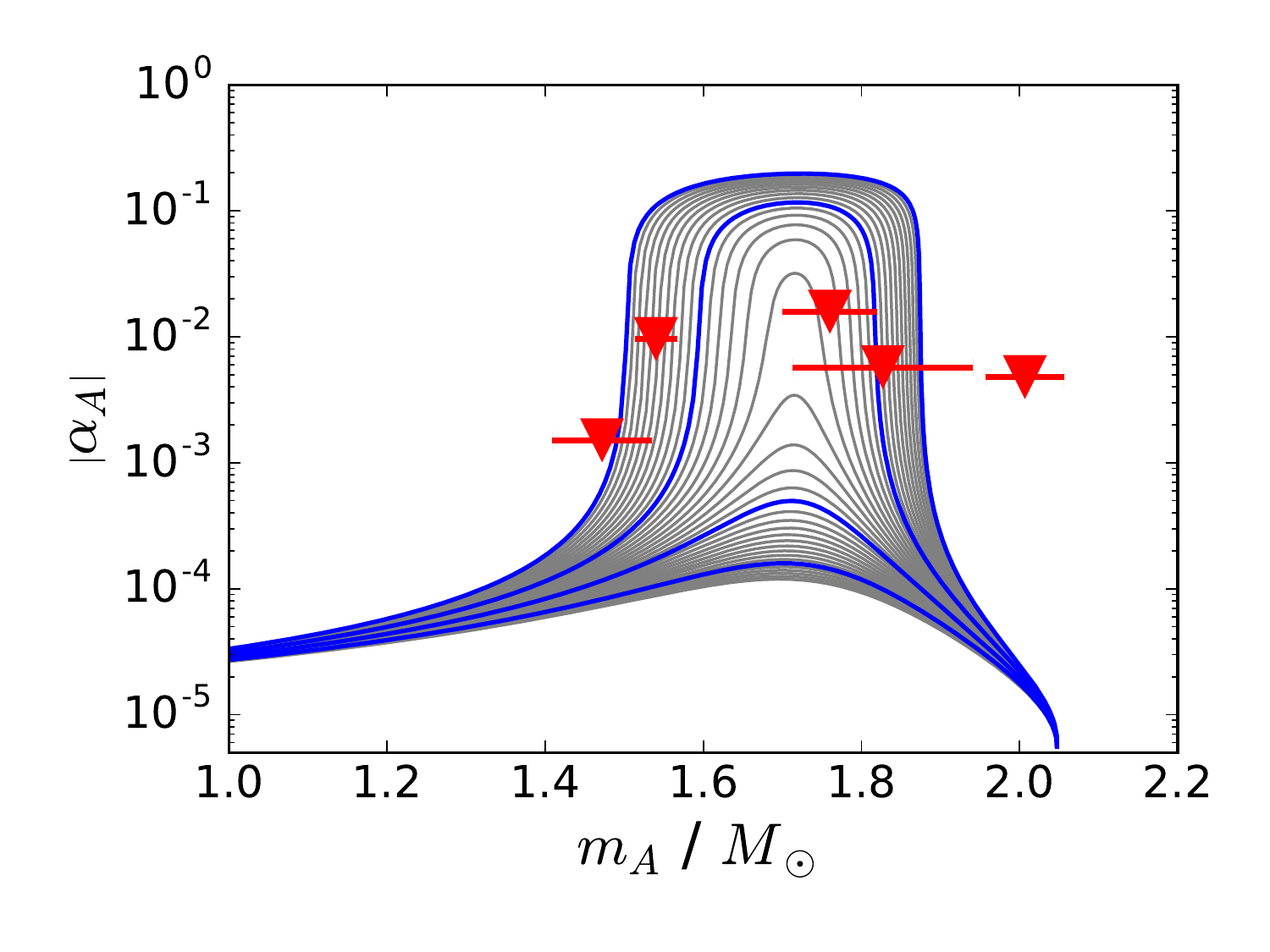}
  \caption{Illustration of spontaneous scalarization in the DEF gravity, in
    comparison to {\it individual} binary-pulsar limits, for a NS with EOS {\sf
    SLy4} and $|\alpha_0| = 10^{-5}$.  The blue curves correspond to (from top
    to bottom) $\beta_0 = -4.5, -4.4, -4.3$, and $-4.2$; the grey curves in
    between differ in $\beta_0$ in steps of $0.01$. We indicate with triangles
    the 90\% CL upper limits on the effective scalar coupling
    $\left|\alpha_A\right|$ from the individual pulsars listed in
    Table~\ref{tab:psr:par}. We can clearly see a scalarization window at $m_A
    \sim 1.7\,M_\odot$.
  \label{fig:ss:SLy4}}
\end{figure}

Using a perfect-fluid description of the energy-momentum tensor for NSs in the
Jordan frame, in 1993 Damour and Esposito-Far{\`e}se derived the
Tolman-Oppenheimer-Volkoff (TOV) equations~\cite{Damour:1993hw} for a NS in
their scalar-tensor gravity theory.  Interestingly, they discovered a {\it
phase-transition} phenomenon when $\beta_0 \lesssim -4$, largely irrespective
of the $\alpha_0$ value (a nonzero $\alpha_0$ tends to smooth the phase
transition~\cite{Damour:1996ke}). The phenomenon was named {\it spontaneous
scalarization}. With a suitable $(\alpha_0,\beta_0)$, the ``effective scalar
coupling'' that a NS develops, $\alpha_A \equiv \partial \ln m_A / \partial
\varphi_0$ (the baryonic mass of NS is fixed while taking the derivative),
could be ${\cal O}(1)$ when the NS mass, $m_A$, is within a certain
EOS-dependent range~\footnote{For black holes, the effective scalar coupling
equals to zero. Therefore, the tests performed with binary black
holes~\cite{TheLIGOScientific:2016src} do not directly apply to the DEF
theory.}.  For masses below and above this range, the effective scalar coupling
is much smaller~\footnote{For sufficiently negative $\beta_0$ ($\lesssim
-4.6$), NSs do not de-scalarize before reaching their maximum mass, i.e.\
spontaneous scalarization is found for all NSs above a certain critical mass,
which depends on the actual value of $\beta_0$ and the EOS~\cite{Damour:1993hw,
Damour:1996ke}.}.  In Fig.~\ref{fig:ss:SLy4} we show an example of spontaneous
scalarization for a NS with the realistic EOS {\sf SLy4}, and compare it to
existing {\it individual} binary-pulsar constraints.

In general, if two compact bodies in a binary have effective scalar couplings,
$\alpha_A$ and $\alpha_B$, they produce gravitational dipolar radiation
$\propto (\Delta\alpha)^2$, with $\Delta\alpha \equiv \alpha_A - \alpha_B$,
which is at a lower post-Newtonian (PN) order than the canonical quadrupolar
radiation in GR~\cite{Damour:1996ke}~\footnote{In this paper, generally we
denote with $n$PN the ${\cal O}(v^{2n}/c^{2n})$ corrections to the leading
Newtonian dynamics (equations of motion).  Therefore, the gravitational dipolar
radiation reaction is at $1.5$\,PN, and the quadrupolar radiation is at
$2.5$\,PN. In the GW phasing, when there is no potential confusion we sometimes
refer to the quadrupolar (dipolar) radiation as $0$\,PN ($-1$\,PN), as
typically done in the literature.}.  In Ref.~\cite{Damour:1998jk}, Damour and
Esposito-Far{\`e}se for the first time compared limits on the DEF gravity
arising from Solar system and binary pulsar experiments with expected limits
from ground-based GW detectors like LIGO and Virgo. The analysis in
Ref.~\cite{Damour:1998jk} is based on soft (by now
excluded~\cite{Demorest:2010bx, Antoniadis:2013pzd}), medium and stiff EOSs,
and for the LIGO/Virgo experiment it assumes a BNS merger with PSR~B1913+16
like masses (1.44\,$M_\odot$ and 1.39\,$M_\odot$), as well as a
1.4\,$M_\odot$-10\,$M_\odot$ NS-BH merger.  Damour and Esposito-Far{\`e}se come
to the conclusion that binary-pulsar experiments would generally be expected to
put more stringent constraints on the parameters $(\alpha_0,\beta_0)$ than
ground-based detectors, such as LIGO and Virgo. Since then, several analyses
have followed~\cite{EspositoFarese:2004tu, EspositoFarese:2004cc,
Freire:2012mg, Antoniadis:2013pzd, Wex:2014nva}, but typically those studies
did not probe a large set of NS masses and EOSs.  Considering advances in our
knowledge of NSs and more sensitive current and future ground-based detectors,
we revisit this study here. Quite interestingly, as pointed out in a first
study in Ref.~\cite{Shibata:2013pra}, the constraints on the parameters
$(\alpha_0, \beta_0)$ from binary pulsars depend quite crucially on the EOSs
and the masses of the NSs, in particular in the parameter space that allows for
spontaneous scalarization. By taking into account this dependence when setting
bounds from pulsar timing, we shall find that current and future GW detectors
on the Earth might still be able to exclude certain specific regions of the
parameter space $(\alpha_0,\beta_0)$ that are not probed by binary pulsars yet.

Twenty years after the discovery of spontaneous scalarization, Barausse {\it et
al.}~\cite{Barausse:2012da} found another interesting nonperturbative
phenomenon in a certain parameter space of the DEF theory. This time the
scalarization does not take place for a NS in isolation, but for NSs in a
merging binary. Indeed, modeling the BNS evolution in numerical relativity,
Barausse {\it et al.} found that the two NSs can scalarize even if initially,
at large separation, they are not scalarized.  This phenomenon is called {\it
dynamical scalarization}, and its onset is determined by the binary compactness
instead of the NS compactness. Reference~\cite{Barausse:2012da} also
demonstrated that a spontaneously scalarized NS can generate scalar hair on its
initially unscalarized NS companion in a binary system through a process known
as \textit{induced scalarization}. Dynamical and induced scalarization cause
BNSs to merge earlier~\cite{Barausse:2012da, Shibata:2013pra,
Taniguchi:2014fqa} than in GR, resulting in a significant modification to the
GW phasing that is potentially detectable by ground-based GW
interferometers~\cite{Barausse:2012da, Palenzuela:2013hsa, Sampson:2014qqa,
Sennett:2016rwa}.

Finally, it is important to note that cosmological solutions in the strictly
massless limit of the DEF theory are known to evolve away from GR when
$\beta_0$ is negative~\cite{Damour:1992kf, Damour:1993id, Sampson:2014qqa,
Anderson:2016aoi}; to be consistent with current Solar-system observations,
such cosmologies require significant fine tuning of initial
conditions.\footnote{For cosmologies in the scalar-tensor theories with a
positive $\beta_0$, we refer readers to Refs.~\cite{Damour:1992kf,
Damour:1993id, Anderson:2017phb}.} Various modifications to the theory have
been considered to cure this fine-tuning problem, e.g. adding higher order
polynomial terms to $\ln A(\varphi)$~\cite{Anderson:2016aoi} or including a
mass term $V(\varphi)=2 m_\varphi \varphi^2$ in the
action~\cite{Damour:1996ke,Ramazanoglu:2016kul, Yazadjiev:2016pcb,
Alby:2017dzl}. To date, none of these proposals have produced scalar-tensor
theories that: (i) satisfy cosmological and weak-field gravity constraints,
(ii) generate an asymptotic field $\varphi_0$ that is stable over timescales
relevant to binary pulsars and GW sources, and (iii) give rise to the
nonperturbative phenomena present in DEF theory. As is commonly done in the
literature~\cite{Damour:1992we, Damour:1993hw, Damour:1996ke, Barausse:2012da,
Shibata:2013pra, Sampson:2014qqa, Palenzuela:2013hsa, Taniguchi:2014fqa,
Sennett:2016rwa}, we will ignore these cosmological concerns in this paper and
focus only on (massless) DEF theory.

%-------------------------------------------------------------------------------
\begin{table*}
  \caption{Binary parameters of the five NS-WD systems that we use to constrain
  the DEF theory~\cite{Antoniadis:2013pzd, Lazaridis:2009kq, Freire:2012mg,
  Reardon:2015kba, Cognard:2017xyr}. The observed time derivatives of the orbit
  period $P_b$ are corrected using the latest Galactic potential of
  Ref.~\cite{McMillan2017}.  For PSRs~J0348+0432, J1012+5307 and J1738+0333,
  the mass ratios were obtained combining radio timing and optical
  high-resolution spectroscopy, while the companion masses are determined from
  the Balmer lines of the WD spectra based on WD
  models~\cite{Antoniadis:2013pzd, Callanan1998, Antoniadis:2016hxz,
  Antoniadis:2012vy}. For PSRs~J1909$-$3744 and J2222$-$0137, the masses were
  calculated from the Shapiro delay, where the range of the Shapiro delay gives
  directly the companion mass, and the pulsar mass is then being derived from
  the mass function, using the shape of the observed Shapiro delay to determine
  the orbital inclination~\cite{Reardon:2015kba, Cognard:2017xyr}. The masses
  below are based on GR as the underlying gravity theory. However, since the
  companion WD is a weakly self-gravitating body, they are practically the same
  in the DEF theory (with a difference $\lesssim 10^{-4}$).  We give in
  parentheses the standard 1-$\sigma$ errors in units of the least significant
  digit(s).
    \label{tab:psr:par}}
  \centering
  \def\arraystretch{1.3}
  \begin{tabularx}{\textwidth}{llllll}
    \hline\hline
    Pulsar & J0348+0432~\cite{Antoniadis:2013pzd} &
    J1012+5307~\cite{Lazaridis:2009kq} & J1738+0333~\cite{Freire:2012mg} &
    J1909$-$3744~\cite{Reardon:2015kba} & J2222$-$0137~\cite{Cognard:2017xyr} \\
    \hline
    Orbital period, $P_b$\,(d) & 0.102424062722(7) & 0.60467271355(3) &
    0.3547907398724(13) & 1.533449474406(13) & 2.44576454(18) \\
    Eccentricity, $e$ & $2.6(9) \times 10^{-6}$ & $1.2(3) \times 10^{-6}$ &
    $3.4(11) \times 10^{-7}$ & $1.14(10) \times 10^{-7}$ & 0.00038096(4) \\
    Observed $\dot P_b$, $\dot P^{\rm obs}_b$\,(${\rm fs\,s}^{-1}$) & $-273(45)$
    & $-50(14)$ & $-17.0(31)$ & $-503(6)$ & $200(90)$ \\
    Intrinsic $\dot P_b$, $\dot P^{\rm int}_b$\,(${\rm fs\,s}^{-1}$) &
    $-274(45)$ & $-5(9)$ & $-27.72(64)$ & $-6(15)$ & $-60(90)$ \\
    Mass ratio, $q\equiv m_p / m_c$ & 11.70(13) & 10.5(5) & 8.1(2) & \ldots &
    \ldots \\
    Pulsar mass, $m_p^{\rm obs}$\,($M_\odot$) & \ldots & \ldots & \ldots &
    1.540(27) & 1.76(6) \\
    WD mass, $m_c^{\rm obs}$\,($M_\odot$) & $0.1715^{+0.0045}_{-0.0030}$ &
    0.174(7) & $0.1817^{+0.0073}_{-0.0054}$ & 0.2130(24) & $1.293(25)$ \\
    \hline
  \end{tabularx}
\end{table*}
%-------------------------------------------------------------------------------

%-------------------------------------------------------------------------------
\section{Constraints from binary pulsars}
\label{sec:bp}
%-------------------------------------------------------------------------------

\begin{figure}
  \includegraphics[width=9cm]{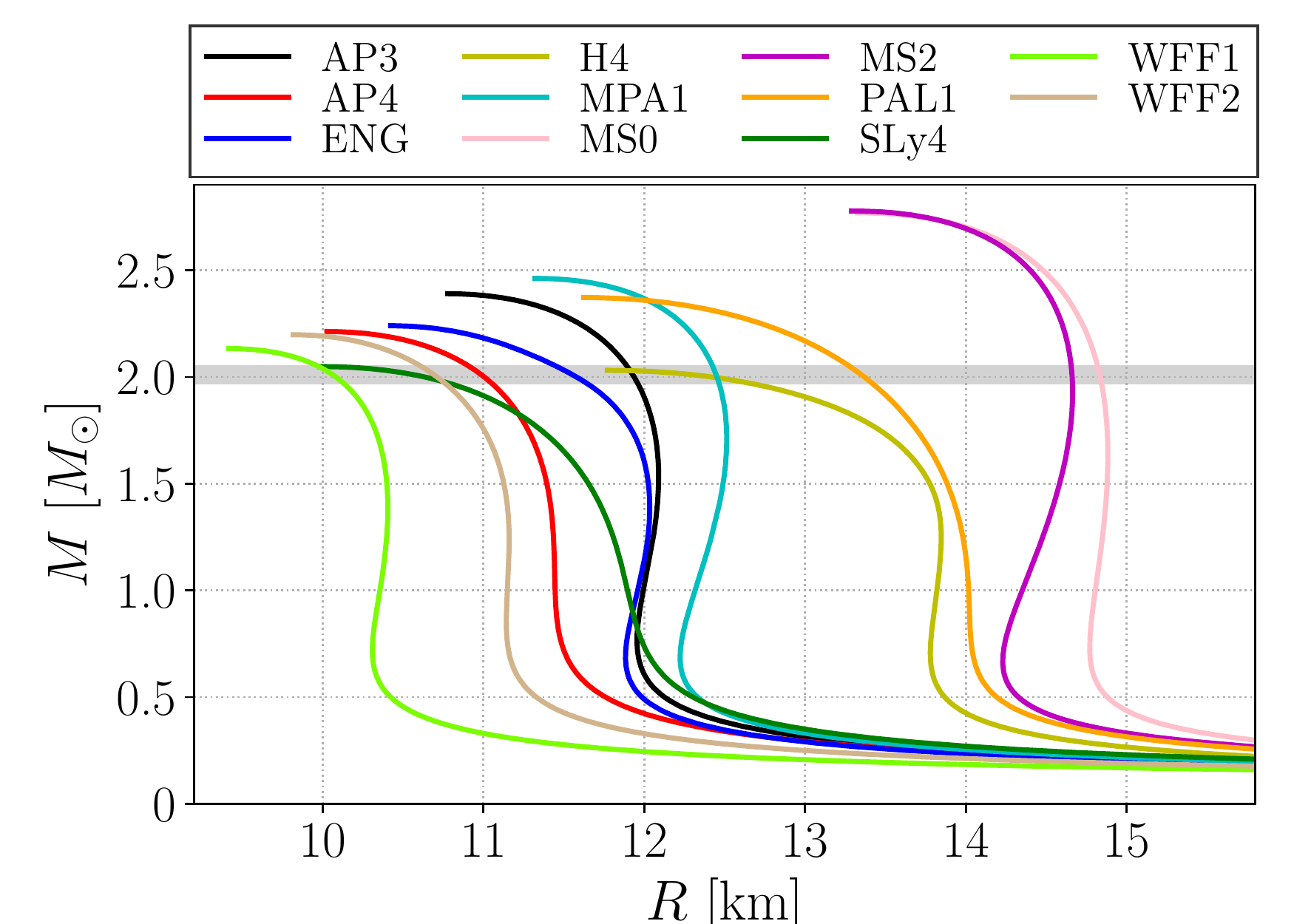}
  \caption{The mass-radius relation of NSs (in GR) for the 11 EOSs that are
    adopted in the study. The mass constraint (with 1-$\sigma$ uncertainty)
    from PSR~J0348+0432~\cite{Antoniadis:2013pzd} is depicted in grey. The
    color coding for different EOSs is kept consistent for all figures in this
    paper.
    \label{fig:eos:mr}}
\end{figure}

Until now, binary pulsars have provided the most stringent limits to the DEF
theory~\cite{Damour:1996ke, Freire:2012mg, Antoniadis:2013pzd, Wex:2014nva,
Cognard:2017xyr}.  These limits were usually obtained with individual pulsar
systems and with representative EOSs~\cite{Wex:2014nva}~\footnote{An exception
  is Ref.~\cite{Kramer:2016kwa}, where, for individual binary pulsars, the most
conservative limits in the $(\alpha_0,\beta_0)$ parameter space across
different EOSs are presented.}.  Here, by contrast, we combine observational
results from multiple pulsar systems employing Markov-chain Monte Carlo (MCMC)
simulations~\cite{DelPozzo:2016ugt}.  In particular, we pick the five NS-WD
binaries that are the most constraining systems in testing spontaneous
scalarization:  PSRs~J0348+0432~\cite{Antoniadis:2013pzd},
J1012+5307~\cite{Lazaridis:2009kq}, J1738+0333~\cite{Freire:2012mg},
J1909$-$3744~\cite{Reardon:2015kba}, and J2222$-$0137~\cite{Cognard:2017xyr}.
We choose these five binaries basing on the binary nature (namely, NS-WD
binaries), the timing precision that has been achieved, and the NS masses.
These aspects are important to the study here, and see
Refs.~\cite{Berti:2015itd, Shao:2016ezh, Kramer:2016kwa} for more discussion.
For convenience, we list the parameters of these binaries  in
Table~\ref{tab:psr:par}, and notice that it is the combination of their
$\dot{P}_b^{\rm int}$ and the NS mass that makes them particularly suitable for
the test of spontaneous scalarization. We obtain the limits using 11 different
EOSs that have the maximum NS mass above $2\,M_\odot$~\cite{Lattimer:2012nd}.
The names of these EOSs are {\sf AP3}, {\sf AP4}, {\sf ENG}, {\sf H4}, {\sf
MPA1}, {\sf MS0}, {\sf MS2}, {\sf PAL1}, {\sf SLy4}, {\sf WFF1}, and {\sf WFF2}
(see Refs.~\cite{Lattimer:2000nx, Lattimer:2012nd} for reviews).
Figure~\ref{fig:eos:mr} shows the mass-radius relation of NSs in GR for these
EOSs. As evidenced by the spread of the curves in the figure, we believe that
these EOSs are sufficient to cover the different EOS-dependent properties of
spontaneous scalarization, and at the same time satisfy the two-solar-mass
limit from pulsar-timing observations~\cite{Demorest:2010bx,
Antoniadis:2013pzd}.

Markov-chain Monte Carlo techniques allow us to preform parameter estimation
within the Bayesian framework. These methods provide the posterior
distributions of the underlying parameters that are consistent with
observations.  In Bayesian analysis, given data ${\cal D}$ and a hypothesis
${\cal H}$ (here, the DEF theory), the marginalized posterior distribution of
$(\alpha_0, \beta_0)$ is given by~\cite{DelPozzo:2016ugt},
\begin{widetext}
\begin{equation}
  P\left(\alpha_0, \beta_0 | {\cal D}, {\cal H}, {\cal I} \right) = \int
  \frac{P\left( {\cal D} | \alpha_0, \beta_0, \bm{\Xi}, {\cal H}, {\cal I}
  \right) P\left( \alpha_0, \beta_0, \bm{\Xi} | {\cal H},{\cal
  I}\right)}{P\left({\cal D} | {\cal H}, {\cal I} \right)} {\rm d} \bm{\Xi} \,,
  \label{eq:bayes}
\end{equation}
\end{widetext}
where ${\cal I}$ is all other relevant prior background knowledge and
$\bm{\Xi}$ collectively denotes all other unknown parameters besides
$(\alpha_0, \beta_0)$, which are marginalized over to obtain the marginalized
posterior distributions for just $(\alpha_0, \beta_0)$ [see below for more
details]. In the above equation, given ${\cal H}$ and ${\cal I}$, $P\left(
\alpha_0, \beta_0, \bm{\Xi} | {\cal H},{\cal I}\right)$ is the prior on
$\left(\alpha_0, \beta_0, \bm{\Xi}\right)$, $P\left( {\cal D} | \alpha_0,
\beta_0, \bm{\Xi}, {\cal H}, {\cal I} \right) \equiv {\cal L}$ is the
likelihood, and $P\left({\cal D} | {\cal H}, {\cal I} \right)$ is the model
evidence.  As said, we use MCMC techniques to explore the posterior in
Eq.~\eqref{eq:bayes}.  We discuss below our choices for the priors  and the
likelihood function [see Eq.~\eqref{eq:log:like}]. We assume that observations
with different binary pulsars are independent.

We now explain how we employ the MCMC technique to get the posterior by
combining binary pulsar systems.  Let us assume that $N$ pulsars ($N = 1,2, 5$,
see below) are used to constrain the $(\alpha_0, \beta_0)$ parameter space. To
obtain a complete description of the gravitational dipolar radiation of these
systems in the DEF theory, we need $N+2$ free parameters in the MCMC runs,
which are $\bm{\theta} = \left\{ \alpha_0, \beta_0, \tilde\rho_c^{(i)}
\right\}$, where $\tilde\rho_c^{(i)}$ ($i=1,\cdots,N$) is the Jordan-frame
central matter density of pulsar $i$~\footnote{Actually, in the full
  calculation we need some other quantities, as well, for example, the orbital
  period, $P_b$, and the orbital eccentricity, $e$. Those quantities are
  observationally very well determined (see Table~\ref{tab:psr:par}), thus we
  use their central values and find that our constraints on
  $(\alpha_0,\beta_0)$ do not change on a relevant scale when we take into
account the errors on those quantities.}. As an initial value to the TOV
solver, we also need the value of the scalar field in the center of a NS,
$\varphi^{(i)}_c$, but the latter is fixed iteratively by requiring that all
pulsars have a common asymptotic value of $\varphi$, $\varphi_0 \equiv \alpha_0
/ \beta_0$.  Given $\tilde\rho_c^{(i)}$ and $\varphi^{(i)}_c$ for pulsar $i$,
we integrate the modified TOV equations (see Eq.~(7) in
Ref.~\cite{Damour:1993hw} or Eq.~(3.6) in Ref.~\cite{Damour:1996ke}) with
initial conditions given by Eq.~(3.14) in Ref.~\cite{Damour:1996ke}. During the
integration, we use tabulated data of EOSs, and linearly interpolate them in
the logarithmic space of the matter density, $\tilde\rho$, the pressure,
$\tilde p$, and the number density, $\tilde n$~\cite{Lattimer:2012nd}.  Note
that only one quantity among $\left\{ \tilde\rho, \tilde p, \tilde n \right\}$
is free, while the others are determined by the EOS. The end products of the
integration provide us, for each pulsar, the gravitational mass, $m_A^{(i)}$,
the baryonic mass, $\bar m_A^{(i)}$, the NS radius, $R^{(i)}$, and the
effective scalar coupling, $\alpha_A^{(i)}$~\cite{Damour:1996ke}.

For the MCMC runs we use a uniform prior on $\log_{10} \left| \alpha_0 \right|$
for $\left|\alpha_0\right| \in [5\times10^{-6}, 3.4\times10^{-3}]$, where
$3.4\times10^{-3}$ is the limit obtained from the Cassini
spacecraft~\cite{Bertotti:2003rm,Damour:2007uf}. We pick the parameter
$\beta_0$ uniformly in the range $[-5, -4]$, which corresponds to a
sufficiently large parameter space where the scalarization phenomena can take
place~\cite{Damour:1993hw,Barausse:2012da}. During the exploration of the
parameter space, we restrict the values of $(\alpha_0, \beta_0)$ to this
rectangle region, as well, in order to avoid overusing computational time in
uninteresting regions. The initial central matter densities, $\left\{
  \tilde\rho_c^{(i)} \right\}$, are picked around their GR values, but they are
  allowed to explore a very large range in the simulations. As we discuss
  below, we perform convergence tests and verify that when evolving the chains
  all parameters in $\bm{\theta}$ quickly lose memory of their initial values.

During the MCMC runs, we evolve the $N+2$ free parameters according to an
affine-invariant ensemble sampler, which was implemented in the {\sf emcee}
package \cite{Goodman:2010,
ForemanMackey:2012ig}~\footnote{\url{http://dan.iel.fm/emcee}}. At every step,
we solve the $N$ sets of modified TOV equations on the fly, using for the
companion  masses of the binary pulsars the values listed in
Table~\ref{tab:psr:par}~\footnote{The masses in PSRs~J0348+0432, J1012+5307,
  and J1738+0333 are based on a combination of radio timing of the pulsars and
  optical spectroscopic observation of the WDs.  The derivation of the masses
  only depends on the well-understood WD atmosphere model in combination with
  gravity at Newtonian order, and the mass ratio $q$, which is free of any
  explicit strong-field effects~\cite{Damour:2007uf}.  Therefore, even within
  the DEF theory, these masses are valid~\cite{Antoniadis:2013pzd,
  Freire:2012mg}.  For PSRs~J1909$-$3744 and J2222$-$0137, the masses are
  derived from the range and shape of the Shapiro delay \cite{Damour:1996ke,
  Wex:2014nva}. Since for the weakly self-gravitating WD companion $|\alpha_B|
  \approx |\alpha_0| \ll 1$, these masses are practically identical to the GR
  masses in Table~\ref{tab:psr:par}.}.

Then, for the decay of the binary's orbital period, which enters the likelihood
function [see Eq.~\eqref{eq:log:like}], we use the dipolar contribution from
the scalar field and the quadrupolar contribution from the tensor field as
given by the following, well known, formulae~\cite{Damour:1992we,
Peters:1964zz},
\begin{equation}
  \dot P_b^{\rm dipole} = - \frac{2 \pi  G_*}{c^3} g(e)
  \left(\frac{2\pi}{P_b}\right)
  \frac{m_p m_c}{m_p + m_c}
  \left(\alpha_A - \alpha_0\right)^2 \,,
  \label{eq:pbdot:dipole}
\end{equation}
\begin{equation}
  \dot P_b^{\rm quad} = -\frac{192\pi G_*^{5/3}}{5c^5}
  f(e) \left(\frac{2\pi}{P_b}\right)^{5/3}
  \frac{m_p m_c}{\left(m_p + m_c\right)^{1/3}} \,,
  \label{eq:pbdot:quad}
\end{equation}
with
\begin{eqnarray}
  g(e) &\equiv& \left(1 + \frac{e^2}{2}\right)\left(1 - e^2\right)^{-5/2} \,, \\
  f(e) &\equiv& \left(1 + \frac{73}{24} e^2 + \frac{37}{96} e^4\right) \left(
  1-e^2\right)^{-7/2} \,.
\end{eqnarray}
We find that higher order terms, as well as the subdominant scalar quadrupolar
radiation, give negligible contributions to this study. Notice that in
Eq.~(\ref{eq:pbdot:dipole}), we have replaced the effective scalar coupling of
the WD companion with the linear matter-scalar coupling constant, since for a
weakly self-gravitating WD $\alpha_A \simeq \alpha_0$ in the $\beta_0$ range of
interest.\footnote{In this context, see footnote ``d'' in
  Ref.~\cite{EspositoFarese:2004tu}, concerning WDs and very large (positive)
$\beta_0$.} Furthermore, we can approximate the bare gravitational constant
$G_*$ in the above equations with the Newtonian gravitational constant $G_N =
G_*(1 + \alpha_0^2)$, since $\left|\alpha_0\right| \ll 1$ ({\it e.g.}, from the
Cassini spacecraft~\cite{Bertotti:2003rm, Damour:2007uf}).

We construct the logarithmic likelihood for the MCMC runs as,
\begin{equation}
  \ln {\cal L} \propto - \frac{1}{2}\sum_{i=1}^N \left[ \left( \frac{ \dot
  P_b^{\rm int} - \dot P_b^{\rm th}}{\sigma_{\dot P_b}^{\rm obs}} \right)^2 +
\left( \frac{m_p/m_c - q}{\sigma_q^{\rm obs}}  \right)^2 \right] \,,
  \label{eq:log:like}
\end{equation}
where for PSRs~J1909$-$3744 and J2222$-$0137 we replace the second term in the
squared brackets with $\left[\left( m_p - m_p^{\rm obs} \right) /
\sigma_{m_p}^{\rm obs}\right]^2$.  In Eq.~(\ref{eq:log:like}), the predicted
orbital decay from the theory is $\dot P_b^{\rm th} \equiv \dot P_b^{\rm
dipole} + \dot P_b^{\rm quad}$, and $\sigma_X^{\rm obs}$ is the observational
uncertainty for $X \in \left\{ \dot P_b^{\rm int}, q, m_p \right\}$, as given
in Table~\ref{tab:psr:par}.  Note that $\dot P_b^{\rm th}$ and $m_p$ implicitly
depend on $(\alpha_0, \beta_0, \bm{\Xi})$, through direct integration of TOV
equations in the DEF theory.

For each EOS, we perform four separate MCMC runs:
\begin{enumerate}[(i)]
\item 1 pulsar: PSR~J0348+0432 ({\sf J0348});
\item 1 pulsar: PSR~J1738+0333 ({\sf J1738});
\item combining 2 pulsars: PSRs~J0348+0432 and J1738+0333 ({\sf 2PSRs});
\item combining 5 pulsars: PSRs~J0348+0432, J1012+5307, J1738+0333,
  J1909$-$3744 and J2222$-$0137 ({\sf 5PSRs}).
\end{enumerate}
We pick {\sf J0348} and {\sf J1738} due to their mass difference
($2.01\,M_\odot$ and $1.46\,M_\odot$ respectively), and their high timing
precision (see Table~\ref{tab:psr:par}), which leads to interesting differences
in the constraints on the DEF parameters, especially on $\beta_0$.  For each
run,  we accumulate sufficient MCMC samples to guarantee the convergence of
MCMC runs. By using the Gelman-Rubin statistic~\cite{Gelman:1992zz}, we find
that, for each EOS, 200,000 samples for cases {\sf J0348} and {\sf J1738}, and
400,000 samples for cases {\sf 2PSRs} and {\sf 5PSRs}, are enough,
respectively. We discard the first half chain points of these 44 runs ($4\,{\rm
cases} \times 11$\,EOSs) as the {\sc burn-in} phase~\cite{Brooks:2011,
ForemanMackey:2012ig}, while we use the remaining samples to do inference on
the parameters $(\alpha_0, \beta_0)$.

\begin{figure}
  \includegraphics[width=9cm]{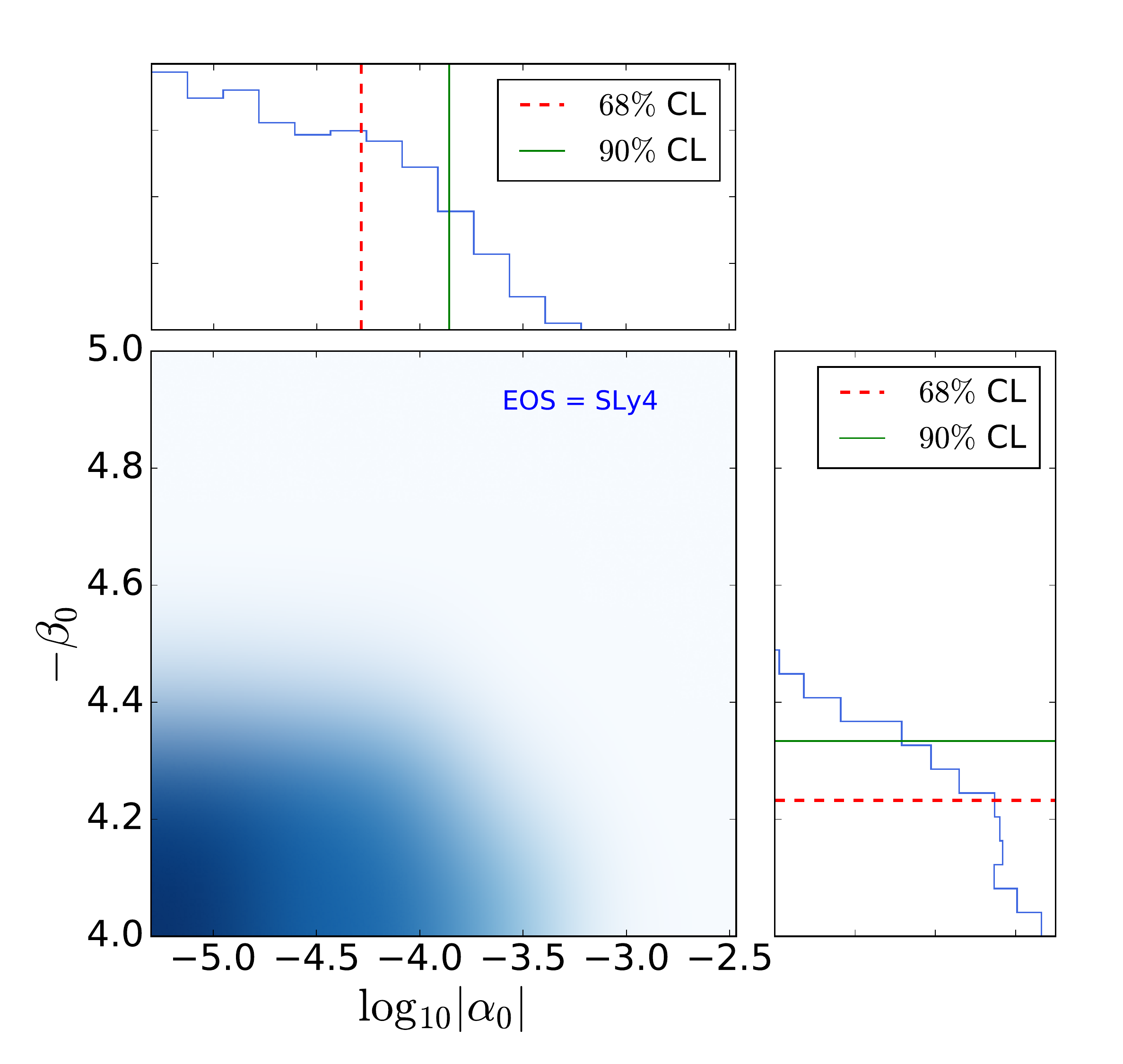}
  \caption{The marginalized 2-d distribution of $(\log_{10}|\alpha_0|,
  -\beta_0)$ from MCMC runs on the five pulsars listed in
  Table~\ref{tab:psr:par}, for the EOS {\sf SLy4}. The marginalized 1-d
  distributions and the extraction of upper limits are illustrated in upper and
  right panels.
  \label{fig:mcmc:SLy4}}
\end{figure}
%%

%-------------------------------------------------------------------------------
\begin{table*}
  \caption{Limits on the parameters of the massless mono-scalar-tensor DEF
    theory for different EOSs when applying the MCMC analysis to the five
    pulsars J0348+0432, J1012+5307, J1738+0333, J1909$-$3744, and J2222$-$0137.
    Results at 68\% and 90\% CLs are listed. $\left| \alpha_A \right|^{\rm
    max}$ is the maximum effective scalar coupling that a NS could still
    possess without violating the limits, and $m_A^{\rm max}$ is the
    corresponding (gravitational) mass at this maximum effective scalar
    coupling (see Figure~\ref{fig:max:alphaA}).
    \label{tab:psr:limit}}
  \centering
  \def\arraystretch{1.2}
  \setlength{\tabcolsep}{0.42cm}
  \begin{tabularx}{\textwidth}{lcccccccc}
    \hline\hline
    & \multicolumn{4}{c}{68\% confidence level} & \multicolumn{4}{c}{90\%
    confidence level} \\
    \hline
    EOS & $\left| \alpha_0 \right|$ & $-\beta_0$ & $m^{\rm max}_A / M_\odot$ &
    $\left| \alpha_A \right|^{\rm max}$ & $\left| \alpha_0 \right|$ &
    $-\beta_0$ & $m_A^{\rm max} / M_\odot$ & $\left| \alpha_A \right|^{\rm
    max}$  \\
    \hline
    {\sf AP3} & $6.5\times10^{-5}$ & $4.21$ & $1.83$ & $1.1\times10^{-3}$
        & $1.5\times10^{-4}$ & $4.29$ & $1.85$ & $6.9\times10^{-3}$ \\
    {\sf AP4} & $5.5\times10^{-5}$ & $4.24$ & $1.71$ & $1.2\times10^{-3}$
        & $1.4\times10^{-4}$ & $4.31$ & $1.73$ & $1.0\times10^{-2}$ \\
    {\sf ENG} & $6.0\times10^{-5}$ & $4.21$ & $1.80$ & $1.0\times10^{-3}$
        & $1.6\times10^{-4}$ & $4.30$ & $1.81$ & $8.2\times10^{-3}$ \\
    {\sf H4}  & $5.7\times10^{-5}$ & $4.24$ & $1.91$ & $1.3\times10^{-3}$
        & $1.7\times10^{-4}$ & $4.33$ & $1.92$ & $2.8\times10^{-2}$ \\
    {\sf MPA1} & $5.7\times10^{-5}$ & $4.22$ & $1.92$ & $1.1\times10^{-3}$
        & $1.6\times10^{-4}$ & $4.30$ & $1.93$ & $8.4\times10^{-3}$\\
    {\sf MS0} & $7.7\times10^{-5}$ & $4.28$ & $2.26$ & $2.7\times10^{-3}$
        & $2.0\times10^{-4}$ & $4.38$ & $2.26$ & $1.0\times10^{-1}$ \\
    {\sf MS2} & $7.9\times10^{-5}$ & $4.26$ & $2.24$ & $2.1\times10^{-3}$
        & $2.4\times10^{-4}$ & $4.36$ & $2.26$ & $8.0\times10^{-2}$ \\
    {\sf PAL1} & $7.3\times10^{-5}$ & $4.21$ & $1.99$ & $1.2\times10^{-3}$
        & $2.0\times10^{-4}$ & $4.29$ & $2.00$ & $8.2\times10^{-3}$ \\
    {\sf SLy4} & $5.2\times10^{-5}$ & $4.23$ & $1.71$ & $1.1\times10^{-3}$
        & $1.4\times10^{-4}$ & $4.33$ & $1.72$ & $2.2\times10^{-2}$ \\
    {\sf WFF1} & $5.3\times10^{-5}$ & $4.21$ & $1.58$ & $9.1\times10^{-4}$
        & $1.3\times10^{-4}$ & $4.30$ & $1.60$ & $6.9\times10^{-3}$ \\
    {\sf WFF2} & $5.5\times10^{-5}$ & $4.24$ & $1.68$ & $1.2\times10^{-3}$
        & $1.4\times10^{-4}$ & $4.32$ & $1.70$ & $1.4\times10^{-2}$ \\
    \hline
  \end{tabularx}
\end{table*}
%-------------------------------------------------------------------------------

As an example, we show in Fig.~\ref{fig:mcmc:SLy4}  the marginalized 2-d
distribution in the parameter space of $(\log_{10}|\alpha_0|, -\beta_0)$ for
the case {\sf 5PSRs} and the EOS {\sf SLy4}.  As mentioned above, we distribute
the initial values of $\log_{10}|\alpha_0|$ and $-\beta_0$ uniformly in the
rectangle region of Fig.~\ref{fig:mcmc:SLy4}.  We see that after MCMC
simulations, the region with large $|\alpha_0|$ or large (negative) $\beta_0$
is no longer populated, and only a small corner is consistent with the
observational results of the five NS-WD binary pulsars.

\begin{figure}
  \includegraphics[width=9cm]{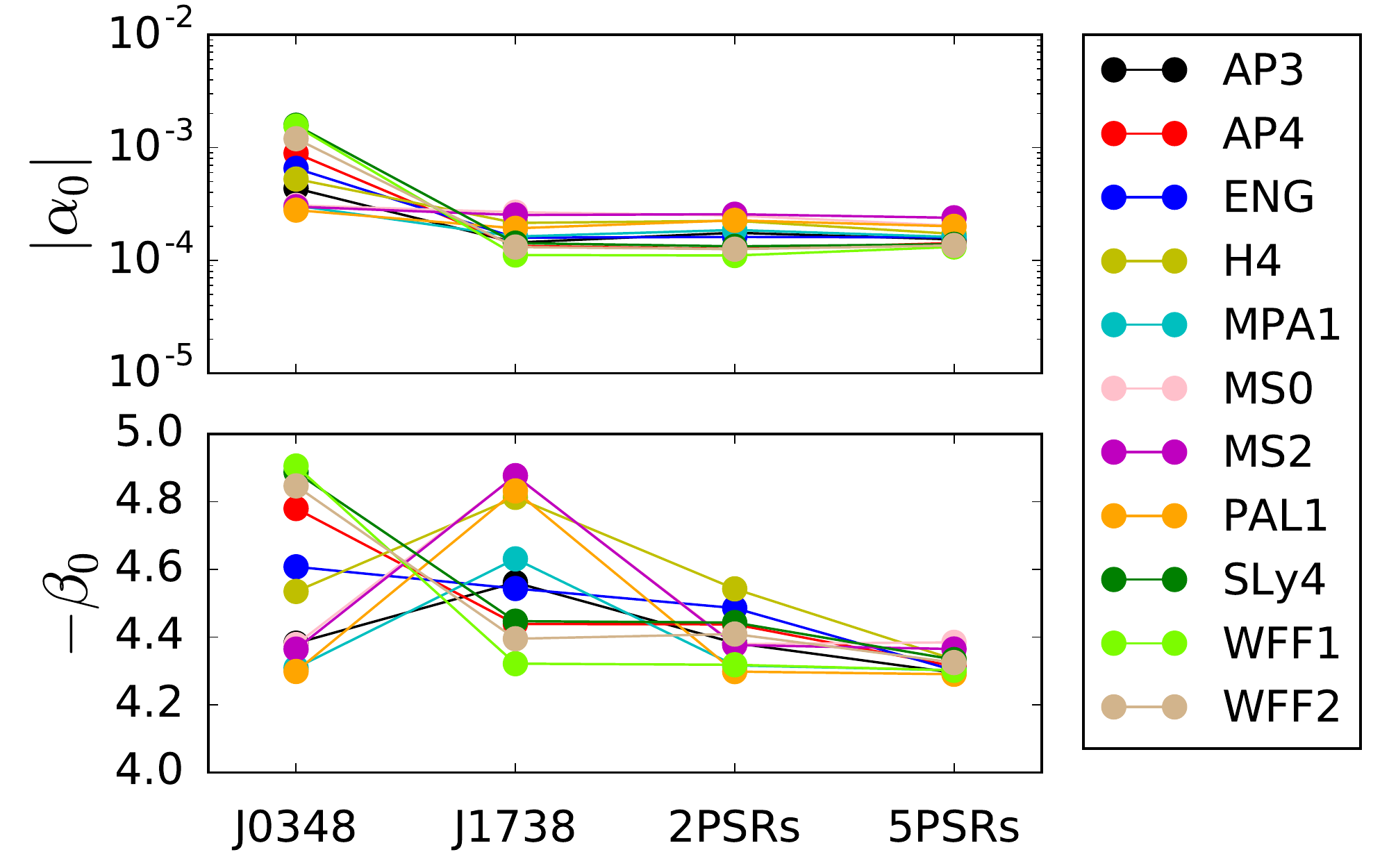}
  \caption{Marginalized upper limits on $|\alpha_0|$ ({\it upper}) and
  $-\beta_0$ ({\it lower}) at 90\%\,CL.  These limits are obtained from PSRs
  J0348+0432 ({\sf J0348}), J1738+0333 ({\sf J1738}), a combination of them
  ({\sf 2PSRs}), and a combination of PSRs~J0348+0432, J1012+5307, J1738+0333,
  J1909$-$3744 and J2222$-$0137 ({\sf 5PSRs}).
  \label{fig:combined:individual:limits}}
\end{figure}

\begin{figure}
  \includegraphics[width=9cm]{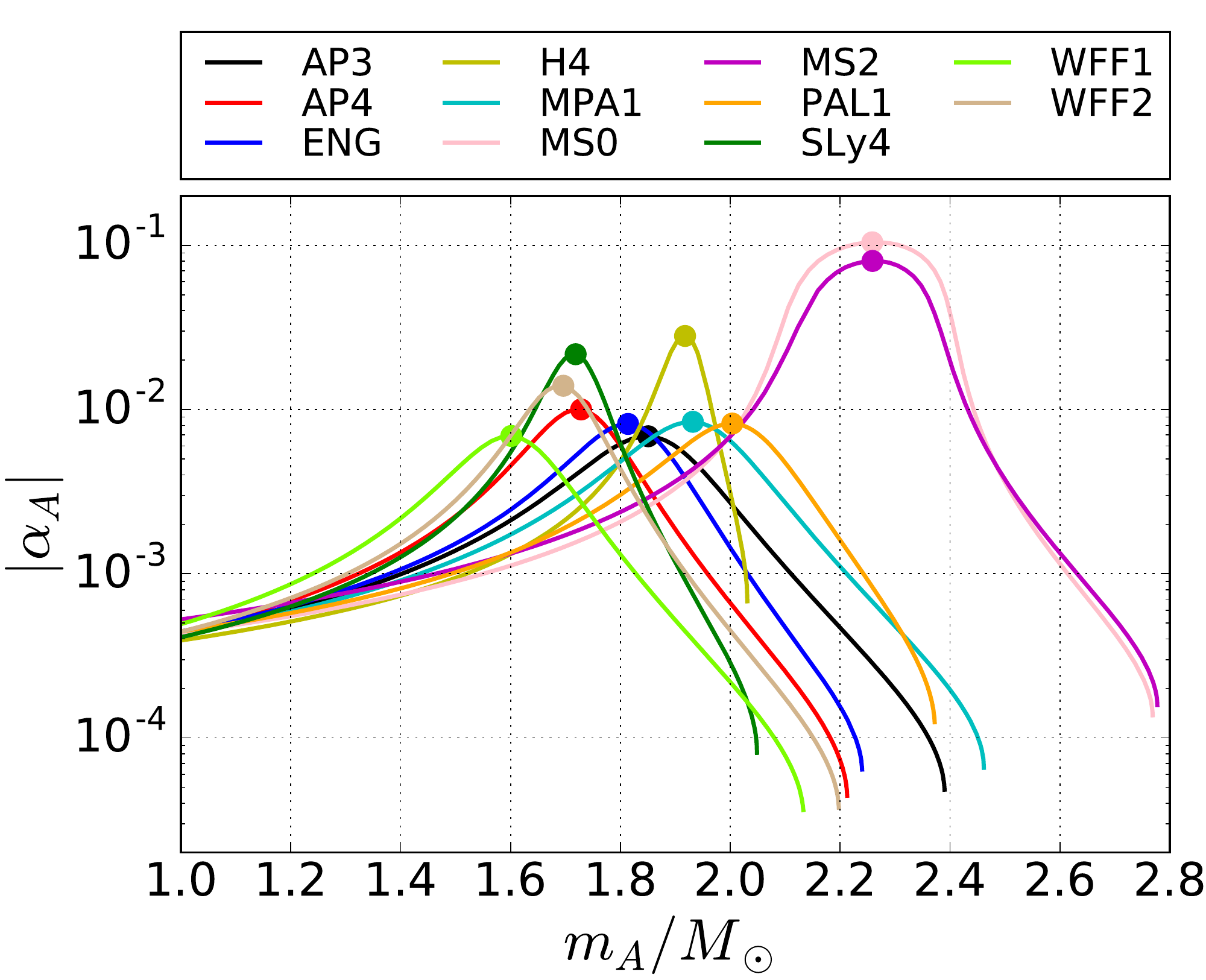}
  \caption{The effective scalar coupling $|\alpha_A|$ that an isolated NS could
    still develop after taking into account the 95\% CL constraints from the
    five pulsars (see Table~\ref{tab:psr:limit}). The point of the maximum
    $|\alpha_A|$ is marked with a dot, and the values (and the corresponding
    masses) are listed in Table~\ref{tab:psr:limit}.
  \label{fig:max:alphaA}}
\end{figure}

Furthermore, we extract the upper limits of $\log_{10}|\alpha_0|$ and
$-\beta_0$ from their marginalized 1-d distributions at 68\% and 90\% CLs. We
summarize the upper limits at 90\% CL from all 44 runs in
Fig.~\ref{fig:combined:individual:limits}. It is interesting to observe the
following facts. First, for all EOSs, {\sf J1738} gives a more constraining
limit on $\alpha_0$ than {\sf J0348}. This result might be due to the fact that
the $\sigma_{\dot P_b}^{\rm obs}$ of {\sf J1738} is about two orders of
magnitude smaller than that of {\sf J0348}, thus giving a tighter limit on
$\alpha_0^2$ by roughly the same order of magnitude. Second, the constraints on
$\beta_0$ from {\sf J0348} and {\sf J1738} are extremely EOS-dependent. This
should be a consequence of the masses of the NSs, which are (in GR)
$1.46\,M_\odot$ for {\sf J1738}, and $2.01\,M_\odot$ for {\sf J0348}. For EOSs
that favour spontaneous scalarization at around $1.4$--$1.5\,M_\odot$, {\sf
J1738} gives a better limit, while for EOSs that favour spontaneous
scalarization at around $2\,M_\odot$, {\sf J0348} gives a better limit. This
trend is also consistent with Fig.~\ref{fig:max:alphaA} (to be introduced
below). Third, by combining two pulsars ({\sf 2PSRs}), NSs are limited to
scalarize at neither $1.4$--$1.5\,M_\odot$ nor $\sim2\,M_\odot$.  Therefore,
almost for all EOSs, $\beta_0$ is well constrained. This result demonstrates
the power of properly using multiple pulsars with different NS masses to
constrain the DEF parameter space for {\it any} EOS. Fourth, we obtain the most
stringent constraints with five pulsars ({\sf 5PSRs}).  This is especially true
for $\beta_0$, which is constrained at the level of $\sim-4.2$ (68\% CL) and
$\sim -4.3$ (90\% CL) for all EOSs.  Finally, we list in
Table~\ref{tab:psr:limit} the marginalized 1-d limits for {\sf 5PSRs}.  We
shall use them in the next section when combining binary pulsars with
laser-interferometer GW observations.

Considering the results that we have obtained when combining the five pulsars
({\sf 5PSRs}), one could wonder whether isolated NSs can still be strongly
scalarized. To address this question, we use the limits on $(\alpha_0,\beta_0)$
and calculate the effective scalar coupling that a NS can still develop as a
function of the NS mass, for the 11 EOSs used in this paper. The results at
90\% CL are summarized in Fig.~\ref{fig:max:alphaA}, while in
Table~\ref{tab:psr:limit} we list the maximally allowed effective scalar
couplings at 68\% and 90\% CLs, and their corresponding (gravitational) NS
masses (marked as dots in Fig.~\ref{fig:max:alphaA}).

Figure~\ref{fig:max:alphaA} clearly shows the {\it nonperturbative} nature of
the scalarization phenomenon. The (absolute values of the) maximally allowed
effective scalar coupling for NSs can be as large as ${\cal O}(10^{-2})$ and
even $0.1$ if the limits at 90\% CL are used, while those values are $\lesssim
10^{-3}$ if one uses the limits at 68\% CL (not shown in
Fig.~\ref{fig:max:alphaA}, but listed in Table~\ref{tab:psr:limit}).
Furthermore, quite remarkably Fig.~\ref{fig:max:alphaA} shows that there are
{\it scalarization windows} (this feature could also be seen in
Fig.~\ref{fig:ss:SLy4} for the EOS {\sf SLy4}). What we mean is the following.
The NS masses for the five most constraining pulsars are $1.46\,M_\odot$
(PSR~J1738+0333) $1.54\,M_\odot$ (PSR~J1909$-$3744), $1.76\,M_\odot$
(PSR~J2222$-$0137), $1.83\,M_\odot$ (PSR~1012+5307), and $2.01\,M_\odot$
(PSR~J0348+0432). For these specific masses, using the 11 EOSs that can give
rise to spontaneous scalarization, we have constrained stringently the DEF
parameters.  However, some EOSs can still allow NS to scalarize strongly ({\it
i.e.}, acquire large effective scalar couplings) for other values of the
masses. As Fig.~\ref{fig:max:alphaA} shows, using limits at 90\% CL, NSs with
EOSs {\sf AP4}, {\sf SLy4}, and {\sf WFF2} can still have $|\alpha_A| \gtrsim
{\cal O}(10^{-2})$ if the NS masses are in the range $m_A =
1.70$--$1.73\,M_\odot$, and NSs with EOS {\sf H4} can still be scalarized to
$|\alpha_A|\sim0.03$ with $m_A \simeq 1.92\,M_\odot$, and NSs with EOSs {\sf
MS0} and {\sf MS2} can still be strongly scalarized to $|\alpha_A| \simeq 0.1$
with $m_A \simeq 2.26\,M_\odot$. Those scalarization windows could be closed in
the future if binary pulsars with these masses are discovered and their
gravitational dipolar radiation is constrained by pulsar timing. As we shall
discuss in Sec.~\ref{sec:gw}, the presence of scalarization windows also opens
the interesting possibility to close these gaps with future GW observations
from BNSs, if the NS's masses lie in the scalarization window.

%===============================================================================
\section{Projected sensitivities for laser-interferometer gravitational-wave
detectors}
\label{sec:gw}
%-------------------------------------------------------------------------------

Having determined constraints on the DEF's parameter space from binary pulsars
(see Table~\ref{tab:psr:limit}) and found scalarization windows, we now address
the question of whether present and future laser-interferometer GW observations
on the ground can still improve these limits and close the gaps.  Two scenarios
are considered: (i) asymmetric BNS systems, equipped with
separation-independent effective scalar couplings, whose gravitational dipolar
radiation during the inspiral can modify the GW
phasing~\cite{Will:1994fb,Damour:1998jk}, and (ii) BNS systems that dynamically
develop scalarization during the late stage of the inspiral, leading to
significant, nonperturbative changes in the GW signal~\cite{Barausse:2012da,
Shibata:2013pra, Taniguchi:2014fqa, Sennett:2016rwa}. A complete description of
BNSs in the DEF theory should include both effects.  However, complete waveform
models from the theory are still not available, so here we investigate the two
scenarios separately to obtain some conservative understanding of the whole
picture.

%-------------------------------------------------------------------------------

\subsection{Dipole radiation for binary neutron-star  inspirals}
\label{subsec:fisher}

The presence of a scalar field can significantly modify the inspiral of an
asymmetric BNS system, due to the additional energy radiated off by the scalar
degree of freedom. The most prominent effect is a modification of the phase
evolution in GW signals. For two NSs with effective scalar couplings $\alpha_A$
and $\alpha_B$ respectively, one finds for the evolution of the orbital
frequency, $\Omega$, up to 2.5\,PN order~\cite{Will:1994fb, Maggiore:1900zz,
Sennett:2016klh},
\begin{equation}
  \label{eq:omdot}
  \frac{\dot{\Omega}}{\Omega^2} = \frac{\eta}{1 + \alpha_A\alpha_B}
    \left[ (\Delta\alpha)^2 {\cal V}^3
           + \frac{96}{5} \kappa {\cal V}^5
           + {\cal O}({\cal V}^6) \right] \,,
\end{equation}
where $\Delta\alpha \equiv \alpha_A - \alpha_B$, $\eta \equiv m_A m_B/M^2$, and
the (dimensionless) ``characteristic'' velocity,
\begin{equation}
  \label{eq:V}
  {\cal V} \equiv \left[ G_\ast \left(1 + \alpha_A\alpha_B \right)M\Omega
  \right]^{1/3}/c \,.
\end{equation}
The quantity $\kappa$ is given in Refs.~\cite{Damour:1992we,
Antoniadis:2013pzd}. In GR one has $\alpha_A = \alpha_B = 0$ and $\kappa = 1$.
Note that there is also a subdominant contribution from the scalar quadrupolar
waves at 2.5\,PN order, which however can be absorbed by a $\lesssim1\%$ change
in the mass parameters.  Here we assume that $\alpha_A$ and $\alpha_B$ are
constant during the inspiral, and their values are obtained from isolated NSs.
This assumption is valid as long as the  induced or dynamical scalarization
mechanisms are not triggered.

For an asymmetric compact binary where $\alpha_A \ne \alpha_B$, the most
prominent deviation from the GR phase evolution is determined by the dipole
term in Eq.~(\ref{eq:omdot}), {\it i.e.}, the contribution $\propto{\cal V}^3$.
To leading order, the offset in the number of GW cycles in band until merger
due to the dipole term is given by
\begin{equation}
  \label{eq:DeltaN}
  \Delta{\cal N}_{\rm dipole} \simeq -\frac{25}{21504\,\pi} \,
    \eta^{-1} \, {\cal V}_{\rm in}^{-7} \, (\Delta\alpha)^2 \,,
\end{equation}
\begin{table*}
  \caption{The number of GW cycles in GR, ${\cal N}_{\rm GR}$, for a BNS merger
  with masses $(1.25\,M_\odot,1.7\,M_\odot)$ for frequencies $f > f_{\rm in}$,
  and its change due to the dipole radiation in the DEF's theory, $\Delta{\cal
  N}_{\rm dipole}$, assuming $|\Delta\alpha| \equiv |\alpha_A - \alpha_B|
  \simeq 0.0199$, which comes from the maximally allowed effective scalar
  couplings for the EOS {\sf SLy4} at 90\% CL (see
  Figure~\ref{fig:max:alphaA}).  The limits on the contributions from
  leading-order spin-orbit and spin-spin terms, $\left|\Delta{\cal N}_\beta
  \right|$ and $\left|\Delta{\cal N}_\sigma\right|$, are listed where the
  (dimensionless) spins of the double pulsar (when it merges in 86\,Myr) are
  used. For $\left|\Delta{\cal N}_\beta \right|$ and $\left|\Delta{\cal
  N}_\sigma\right|$, we also give in parentheses when {\it both} NSs are
  spinning at the maximal spin that we have ever observed (in an eclipsing
  binary pulsar J1748$-$2446ad).
  \label{tab:DelN}}
  \centering
  \def\arraystretch{1.2}
  \setlength{\tabcolsep}{0.5cm}
  \begin{tabular}{llllll}
  \hline\hline
  Detector & $f_{\rm in}$ (Hz) & ${\cal N}_{\rm GR}$ & $\Delta{\cal N}_{\rm
  dipole}$ & $\left|\Delta{\cal N}_\beta\right|$ & $\left|\Delta{\cal
  N}_\sigma\right|$ \\
  \hline
  aLIGO & 10 & $1.5\times10^4$ & $-3.7\times10^1$ & $< 0.76$ \quad ($<
  3.5\times10^1$) & $< 1.8\times10^{-6}$ \quad ($< 0.43$) \\
  CE & 5 & $4.8\times10^4$ & $-1.9\times10^2$ & $< 1.2~$ \quad ($<
  5.6\times10^1$) & $< 2.3\times10^{-6}$ \quad ($< 0.55$) \\
  ET & 1 & $7.0\times10^5$ & $-8.1\times10^3$ & $< 3.5~$ \quad ($<
  1.6\times10^2$) & $< 3.9\times10^{-6}$ \quad ($< 0.93$) \\
  \hline
\end{tabular}
\end{table*}
where ${\cal V}_{\rm in}$ corresponds to ${\cal V}$ in Eq.~(\ref{eq:V}) when
the merging system enters the band of the GW detector, {\it i.e.}, when $\Omega
= \pi f_{\rm in}$ (see Refs.~\cite{Will:1994fb, Damour:1998jk} for details). In
the above equation we have used the approximation $\kappa \simeq 1$ and the
fact that ${\cal V}_{\rm in}$ is much smaller than ${\cal V}$ just before
merger.  Within the approximation of Eq.~(\ref{eq:DeltaN}) one can use ${\cal
V}_{\rm in} \simeq (G_{\rm N}M \pi f_{\rm in})^{1/3}/c$, {\it i.e.}, replacing
the effective gravitational constant $ G_\ast \left(1 + \alpha_A\alpha_B
\right)$ in Eq.~(\ref{eq:V}) by the Newtonian gravitational constant $G_{\rm N}
\equiv G_\ast\left( 1 + \alpha_0^2 \right) $ \cite{Damour:1992we,
Damour:1996ke}.  Again, we stress that Eq.~(\ref{eq:DeltaN}) is based on the
assumption that the effective scalar couplings of the two NSs, $\alpha_A$ and
$\alpha_B$, remain unchanged during the inspiral in the detector's sensitive
frequency band.  It therefore neglects the phenomenon of  induced
scalarization, which can occur in a BNS system, when the unscalarized NS is
sufficiently exposed to the scalar field of the scalarized
companion~\cite{Palenzuela:2013hsa}. This can reduce the dipolar radiation
considerably on short ranges if $\alpha_A$ approaches $\alpha_B$, and lead to a
characteristic change in the late phase evolution of the merging BNSs.  Studies
on the dynamically changing effective scalar couplings are performed in the
next subsection.

To obtain a rough understanding of the effects of dipolar radiation, let us
calculate the dephasing from GR by an asymmetric BNS inspiral with $m_A =
1.25\,M_\odot$ and $m_B = 1.7\,M_\odot$. According to
Fig.~\ref{fig:max:alphaA}, at present, binary-pulsar experiments cannot exclude
$|\alpha_A|$ as large as $10^{-2}$--$10^{-1}$ for NSs of a certain mass range,
which depends on the EOS.  For the EOS {\sf SLy4} we find from the
corresponding (dark green) curve in Fig.~\ref{fig:max:alphaA} $|\alpha_A|
\simeq 0.0007$ and $|\alpha_B| \simeq 0.0206$, hence $|\Delta\alpha| \equiv
|\alpha_A - \alpha_B| \simeq 0.0199$.

In our study we consider the Advanced LIGO (aLIGO) detectors at design
sensitivity~\cite{TheLIGOScientific:2014jea}, and future ground-based
detectors, such as the Cosmic Explorer (CE), and the Einstein Telescope (ET).
We use the starting frequencies, $f_{\rm in}=10$\,Hz for aLIGO, $f_{\rm
in}=5$\,Hz for CE, and $f_{\rm in}=1$\,Hz for ET~\cite{Hild:2010id,
Evans:2016mbw, Shoemaker:2010}. In Table~\ref{tab:DelN} we list the number of
GW cycles as predicted by GR, ${\cal N}_{\rm GR}$, and the change in the number
of cycles caused by the existence of a dipole radiation for a BNS signal,
$\Delta{\cal N}_{\rm dipole}$. From Table~\ref{tab:DelN} we can already see
that, given the BNS parameters, current bounds by pulsars still leave room for
significant time-domain phasing modifications in BNS mergers, in particular if
one of the NSs falls into the scalarization window of $\sim 1.7\,M_\odot$ (for
EOSs {\sf AP4}, {\sf SLy4}, and {\sf WFF2}) to $\sim 1.9\,M_\odot$ (for the EOS
{\sf H4}), or if one NS's mass significantly exceeds $2\,M_\odot$ (for EOSs
{\sf MS0} and {\sf MS2}). As reference points, we list in Table~\ref{tab:DelN}
also the changes in the number of GW cycles from spin-orbit and spin-spin
effects.  Indeed, from the leading-order spin-orbit (1.5\,PN) and spin-spin
(2\,PN) contributions to the GW phasing~\cite{Buonanno:2009zt, Berti:2005qd},
one has
  \begin{eqnarray}
    \Delta {\cal N}_\beta &\simeq& \frac{5}{64\pi} \eta^{-1} {\cal V}_{\rm
    in}^{-2} \beta \,, \\
    \Delta {\cal N}_\sigma &\simeq& - \frac{5}{32\pi} \eta^{-1} {\cal V}_{\rm
    in}^{-1} \sigma \,,
  \end{eqnarray}
  where
  \begin{eqnarray}
   \beta &=& \frac{1}{12} \sum_{i=A,B} \left( 113 \frac{m_i^2}{M^2} +
    75\eta \right) \hat{\bm{L}} \cdot \bm{\chi}_i \,,\\
    \sigma &=& \frac{\eta}{48} \left[ -247 \bm{\chi}_A \cdot \bm{\chi}_B +
      721\left( \hat{\bm{L}} \cdot \bm{\chi}_A\right) \left( \hat{\bm{L}} \cdot
      \bm{\chi}_B \right) \right] \,.
  \end{eqnarray}
The (dimensionless) spins of a BNS system, $\bm{\chi}_A$ and $\bm{\chi}_B$, are
likely to be small in magnitude.  The parameters $\beta$ and $\sigma$ are
maximized when two spins are aligned with the direction of the orbital angular
momentum, $\hat{\bm{L}}$.  The limits on $\left|\Delta{\cal N}_\beta\right|$
and $\left|\Delta{\cal N}_\sigma\right|$ are listed in Table~\ref{tab:DelN}
where we have used the (dimensionless) spins of the double pulsar system that
is the only double NS system where two spins are precisely measured. When the
double pulsar evolves to the time of its merger in 86\,Myr from now, one has
$\left|\bm{\chi}_A\right| \simeq 0.014$ and $\left|\bm{\chi}_B\right| \simeq
0.00002$~\cite{Kramer:2006nb}, assuming a canonical moment of inertia
$10^{38}\,{\rm kg\,m}^2$ for NSs. As we can see from Table~\ref{tab:DelN}, if
the spins of BNSs to be discovered by GW detectors are comparable to that of
the double pulsar, the inclusion of spins only affects the number of GW cycles
at percentage level at most. In addition, because ground-based detectors could
observe BNSs from a population different from the one observed with pulsar
timing, we also give  $\left|\Delta{\cal N}_\beta\right|$ and
$\left|\Delta{\cal N}_\sigma\right|$ in Table~\ref{tab:DelN} when the
(dimensionless) spin of the fastest rotating pulsar ever observed,
PSR~J1748$-$2446ad ($P=1.4\,{\rm ms}$)~\cite{Hessels:2006ze}, is used for {\it
both} NSs~\footnote{Notice that PSR~J1748$-$2446ad is {\it not} in a double NS
binary. Its companion is probably a bloated main-sequence star that recycles
the pulsar to a large spin~\cite{Hessels:2006ze}.}. Even in this extreme case
with $|{\bm \chi}_A| = |{\bm \chi}_B| \simeq 0.26$ (assuming a canonical mass
$1.4\,M_\odot$), $\left|\Delta{\cal N}_{\rm dipole} \right|$ is still larger
(or comparable in the case of the Advanced LIGO) than the {\it upper limits} of
$\left|\Delta{\cal N}_\beta\right|$ and $\left|\Delta{\cal N}_\sigma\right|$.

The dephasing quantity, $\Delta{\cal N}_{\rm dipole}$, is nevertheless a crude
indicator for realistic detectability. In reality, one has to consider various
degeneracy between binary parameters, the waveform templates that are used for
detection and parameter estimation, the power spectral density (PSD) of noises
in GW detectors, $S_n(f)$, the signal-to-noise ratio (SNR) of an event $\rho$,
and so on. In order to obtain more quantitative estimates of the constraints on
dipolar radiation that can be expected from GW detectors, one would need to
compute Bayes factors between two alternatives~\cite{DelPozzo:2014cla} or apply
cutting-edge parameter-estimation techniques, for example  the MCMC or nested
sampling~\cite{Veitch:2014wba}.  However, given the limited scope of our
analysis, for simplicity, here, we adopt the Fisher-matrix
approach~\cite{Finn:1992wt, Cutler:1994ys, Will:1994fb, Berti:2004bd}, although
we are aware of the fact that for events with mild SNR ($\rho \sim 10$), the
Fisher matrix can have pitfalls~\cite{Vallisneri:2007ev}. In
Appendix~\ref{sec:app:fisher} we review the main Fisher-matrix tools that we
use.

\begin{figure}
  \includegraphics[width=9cm]{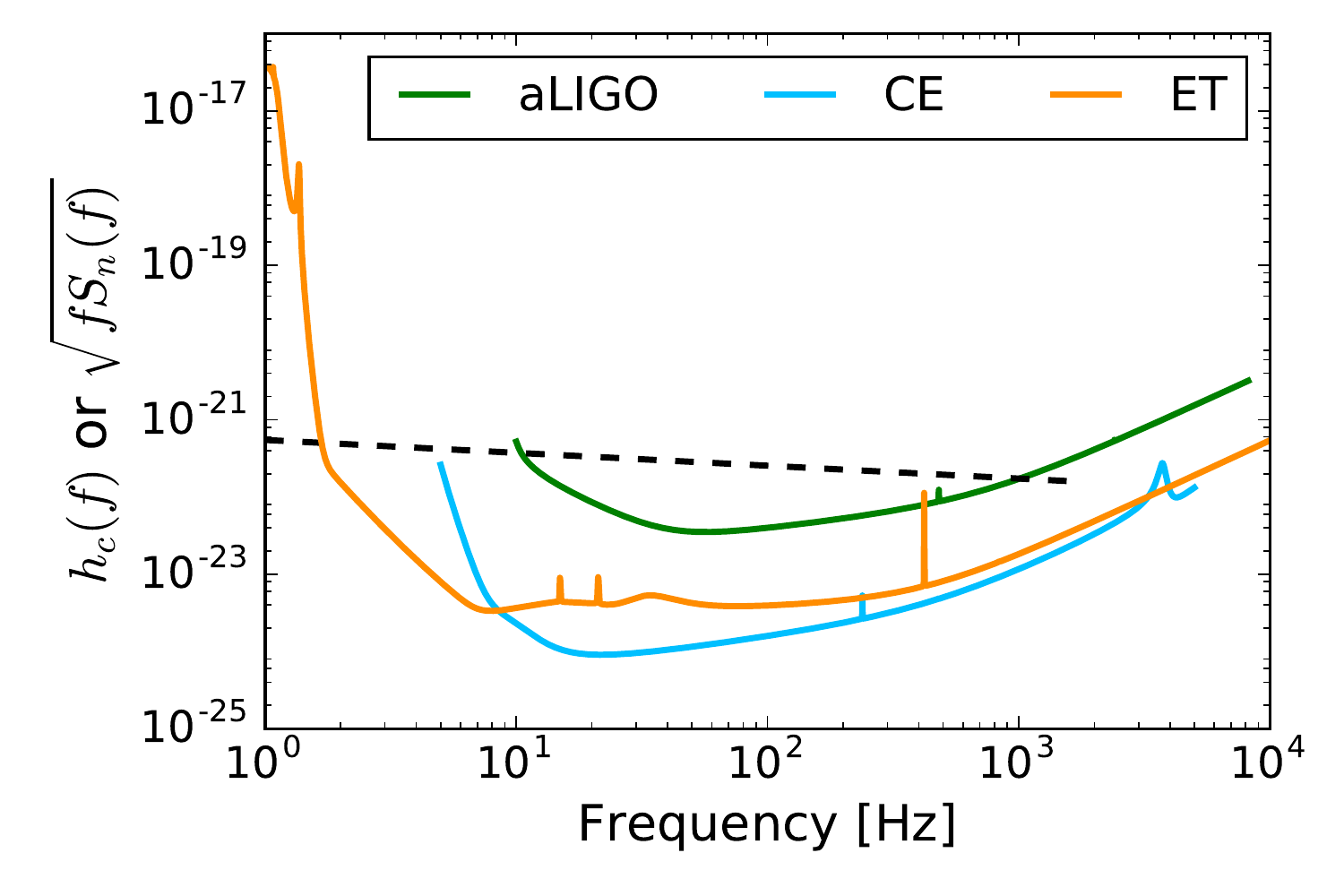}
  \caption{Dimensionless noise spectral density  $\sqrt{fS_n(f)}$ of aLIGO, CE
  and ET GW detectors. The quantity $S_n(f)$ is the one-side design
  PSD~\cite{Hild:2010id, Evans:2016mbw, Shoemaker:2010}. The dashed line
  describes the (dimensionless) pattern-averaged characteristic strain $h_c(f)
  \equiv 2f\left|\tilde h(f) \right|$ for a BNS with rest-frame masses
  $(1.25\,M_\odot,1.63\,M_\odot)$ at 200\,Mpc ($z\simeq 0.0438$), up to the
  innermost-stable circular orbit given by Eq.~(\ref{eq:isco}).
  \label{fig:psd}}
\end{figure}

We summarize in Fig.~\ref{fig:psd} their dimensionless noise spectral density
$\sqrt{fS_n(f)}$~\cite{Hild:2010id, Evans:2016mbw, Shoemaker:2010}, and show in
the figure also an hypothetical BNS signal. For all the studies, we fix the
luminosity distance to $D_L=200$\,Mpc for aLIGO, CE, and ET. Indeed, within
such a distance, aLIGO alone is supposed to observe $0.2$--$200$ BNS events
annually at design sensitivity~\cite{Aasi:2013wya}. With the four-site network
incorporating LIGO-India at design sensitivity, the number of detectable BNS
events will double~\cite{Aasi:2013wya}. Therefore, it is a realistic setting to
discuss BNS events for aLIGO; to be conservative, we only consider a
two-detector network for aLIGO in our study. CE and ET have better
sensitivities, thus will have larger SNRs for these events; besides, they will
be able to detect a larger number of BNSs, including those with unfavourable
orientations. Using the standard cosmological model~\cite{Ade:2015xua}, the
redshift associated to $D_L=200$\,Mpc is $z \simeq 0.0438$, and  we take it
into account in our Fisher-matrix calculation, even if it generates a small
effect. Moreover, we always report masses in the rest frame of a BNS system.

\begin{figure}
  \includegraphics[width=9cm]{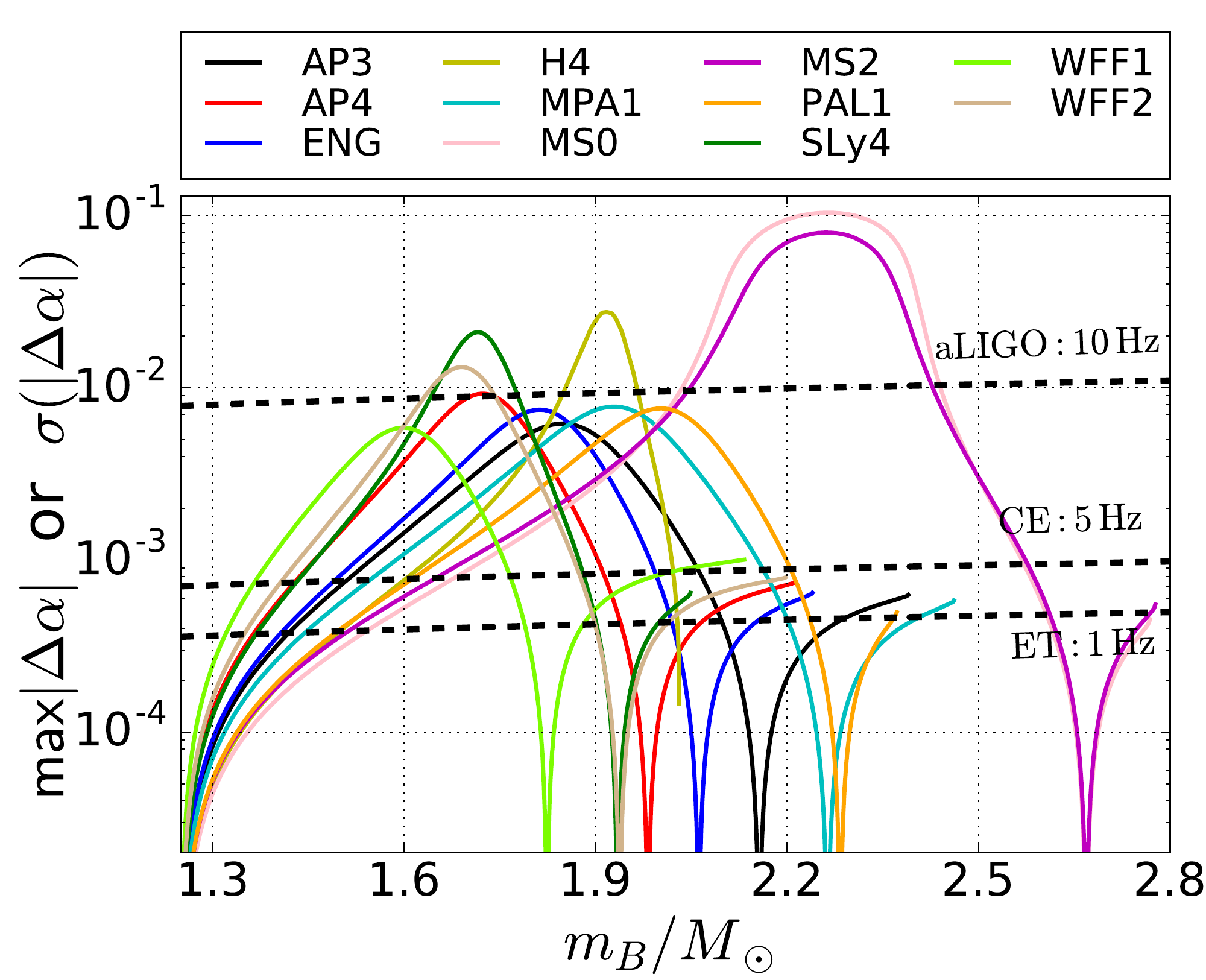}
  \caption{The sensitivities of aLIGO, CE, and ET to
    $\left|\Delta\alpha\right|$ (namely the uncertainty, $\sigma\left(
    |\Delta\alpha| \right)$, obtained from the inverse Fisher matrix) are
    depicted with dashed lines, as a function of $m_B$, for a pattern-averaged
    BNS inspiral signal with rest-frame component masses ($m_A=1.25\,M_\odot$,
    $m_B$). The starting frequencies of GW detectors are labeled. Luminosity
    distance $D_L = 200$\,Mpc is assumed. The sensitivity to
    $\left|\Delta\alpha\right|$ from GW detectors scales with SNR as
    $\rho^{-1/2}$.  The maximum available values of $\left|\Delta\alpha\right|$
    for 11 EOSs, saturating the limits from binary pulsars at 90\% CL, are
    shown in solid lines. If a sensitivity curve (dashed) is below a solid
    curve, the corresponding GW detector has the potential to improve the limit
    from binary pulsars for this particular EOS, with BNSs of suitable masses.
    \label{fig:dipole:inspiral}}
\end{figure}

The Fisher matrix is constructed as usual from the Fourier-domain waveform
$\tilde h(f)$~\cite{Finn:1992wt, Cutler:1994ys, Will:1994fb},
\begin{equation}
  \Gamma_{ab} \equiv \left( \partial_a \tilde h(f) \right. \left| \partial_b
  \tilde h(f) \right) \,,
  \label{eq:fisher:matrix}
\end{equation}
with $\partial_a \tilde h(f)$ being the partial derivative to the parameter
labeled ``$a$'' (see Appendix~\ref{sec:app:fisher} for definitions  and
notations).  We use the waveform parameters $\left\{ \ln{\cal A}, \ln \eta, \ln
{\cal M}, t_c, \Phi_c, \left(\Delta \alpha\right)^2 \right\}$ to construct the
$6\times6$ Fisher matrix, $\Gamma_{ab}$. The inverse of the Fisher matrix is
the correlation matrix for these parameters, from where we can read their
uncertainties and correlations~\cite{Finn:1992wt, Cutler:1994ys, Will:1994fb,
Berti:2004bd}.

\begin{figure*}
  \includegraphics[width=18cm]{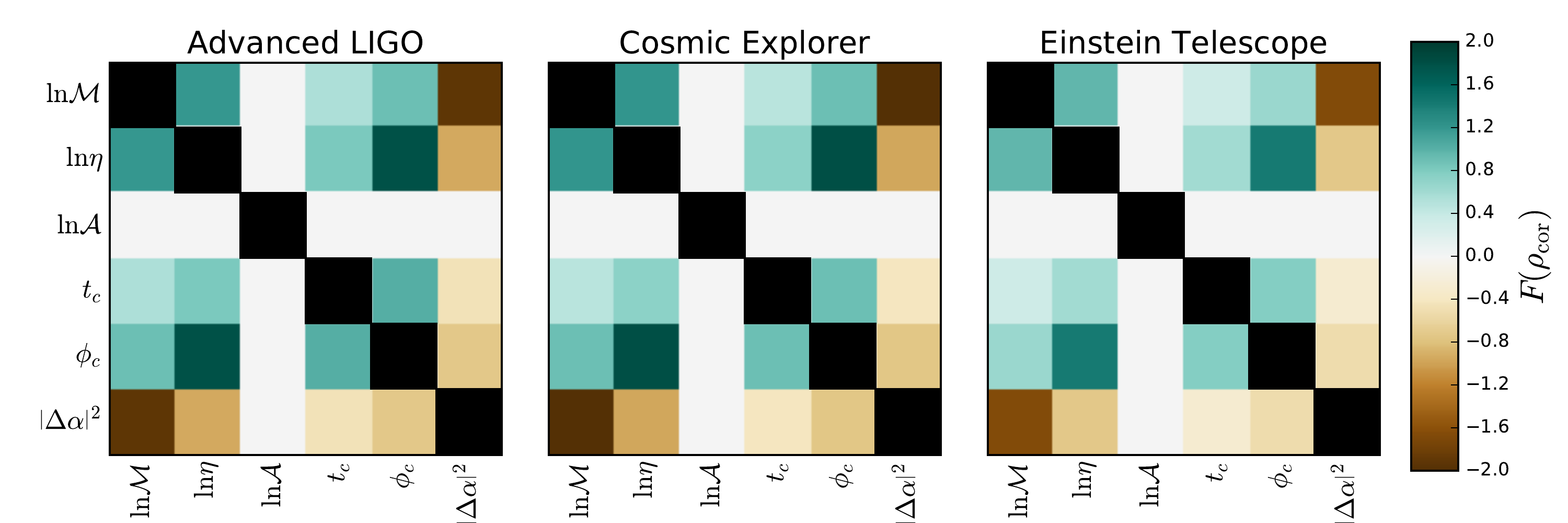}
\caption{The correlations between six parameters, obtained from the inverse
  Fisher matrix in the matched-filter analysis for a BNS with rest-frame masses
  $(1.25\,M_\odot, 1.63\,M_\odot)$. $F(\rho_{\rm cor}) \equiv \log_{10}\left[
  \left( 1+\rho_{\rm cor} \right) / \left( 1-\rho_{\rm cor} \right) \right] -
  \rho_{\rm cor} \log_{10} 2$ is a function defined in Ref.~\cite{Shao:2016ubu}
  such that it {\it counts} 9's in the limit of large correlations [{\it e.g.},
    $F(0.99)\simeq +2$, $F(-0.9)\simeq -1$, and $F(0)=0$]. On diagonal,
    $F(\rho_{\rm cor}=1)$ diverges and is plotted in black.
  \label{fig:corr}}
\end{figure*}

In Fig.~\ref{fig:dipole:inspiral} we plot in dashed lines the uncertainties in
$|\Delta\alpha|$ obtained with three GW detectors (aLIGO, CE, and ET) for an
asymmetric BNS with rest-frame masses $m_A=1.25\,M_\odot$ and $m_B >
1.25\,M_\odot$,  located at $D_L=200$\,Mpc.  For a BNS of masses, for example,
$(1.25\,M_\odot, 1.63\,M_\odot)$ which are the most probable masses for the
newly discovered asymmetric double-NS binary pulsar
PSR~J1913+1102~\cite{Lazarus:2016hfu}, we find that aLIGO, CE, and ET can
detect its merger at 200\,Mpc with $\rho=10.6$, $450$, and $153$, respectively,
after averaging over pattern functions and assuming two detectors in each case.
The characteristic strain of such a BNS is illustrated in the frequency domain
in Fig.~\ref{fig:psd}. In the large SNR limit, the uncertainties in
$|\Delta\alpha|$  scale with the SNR as $\rho^{-1/2}$. In Fig.~\ref{fig:corr},
we give the correlations between parameters obtained from the matched-filter
analysis. We find that due to its low-frequency sensitivity ET can break some
degeneracy between parameters better than aLIGO and CE do.

In Fig.~\ref{fig:dipole:inspiral}, we show with solid lines the maximum values
of $|\Delta\alpha|$ at 90\% CL from pulsars for 11 EOSs (calculated from
Fig.~\ref{fig:max:alphaA}). If for some NS's mass range a solid line (which is
associated to a certain EOS) is above a dashed line (associated to a certain
detector), then for NSs described by that EOS, the corresponding GW detector
has potential to further improve the DEF's parameters with the observation of a
BNS within that mass range.  From the figure we can see that, with the expected
design sensitivity curves of aLIGO, CE, and ET~\cite{Shoemaker:2010,
Hild:2010id, Evans:2016mbw},
\begin{itemize}
  \item aLIGO has potential to further improve the current limits from binary
    pulsars with a discovery of a BNS of suitable masses, if the EOS of NSs is
    one of (or similar to) {\sf H4}, {\sf MS0}, {\sf MS2}, {\sf SLy4}, and {\sf
    WFF2};
  \item CE and ET, due to their low-frequency sensitivity and better PSD
    curves, are able to significantly improve current limits from binary
    pulsars on the DEF's parameters, no matter what the real EOS of NSs is.
\end{itemize}
We stress that those conclusions are obtained with a Fisher-matrix analysis,
and should be made more robust in the future using more sophisticated tools,
notably Bayesian analysis.

%---------------------------------------------------------------------
\subsection*{Constraints outside the spontaneous-scalarization regime}
%---------------------------------------------------------------------

\begin{figure*}
  \includegraphics[width=12cm]{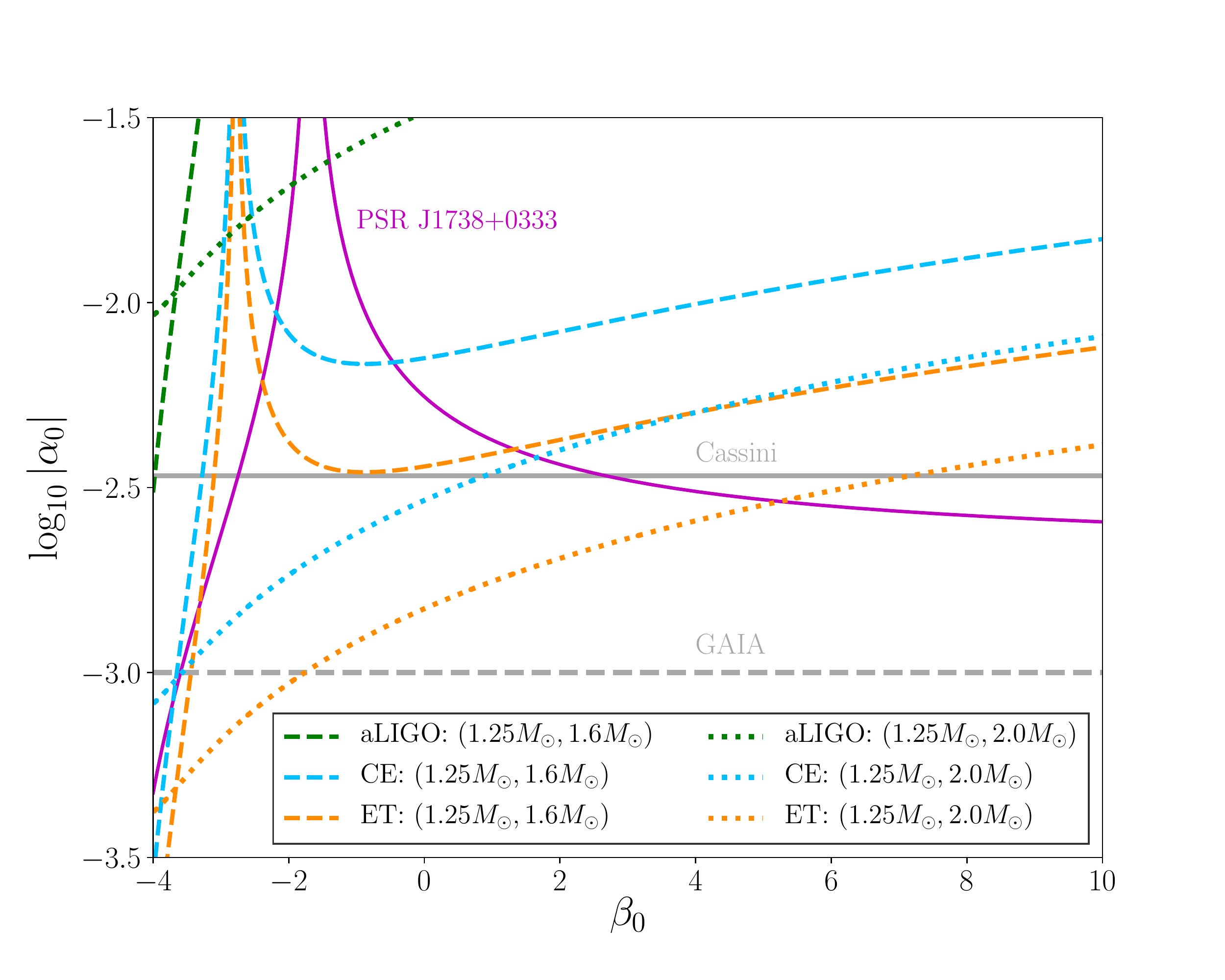}
  \caption{Upper limits on $|\alpha_0|$ as a function of $\beta_0$ for aLIGO
    (green), CE (blue), and ET (orange)~\cite{Damour:1998jk}. The dashed lines
    correspond to a 1.25\,$M_\odot$/1.6\,$M_\odot$ BNS merger, and the dotted
    lines correspond to a 1.25\,$M_\odot$/2.0\,$M_\odot$ BNS merger. The chosen
    EOS is {\sf AP4}.  For comparison we have plotted Solar system limits
    (grey) and the limits from PSR~J1738+0333 (magenta), which currently gives
    the best limit for $\beta_0 \gtrsim 3$. The limit from
    Cassini~\cite{Bertotti:2003rm} and the limit expected from
    GAIA~\cite{2010IAUS..261..306M} are also shown.
    \label{fig:b0gem4}}
\end{figure*}

With the results above it is fairly straightforward to calculate the limits
from aLIGO, CE and ET on $\left| \alpha_0 \right|$ when $\beta_0$ is outside
the spontaneous scalarization regime, {\it i.e.},\ $\beta_0 \gtrsim -4$, and
compare them to existing limits from the Solar system and
pulsars~\cite{Damour:1998jk}.  For completeness, we present the relevant
results here.  \citet{Shibata:2013pra} have shown that for small $\alpha_0$
there exists a simple relation between $\alpha_A$, $\alpha_0$ and $m_A$ as long
as spontaneous scalarization does not set in (see Eq.~(44) in
Ref.~\cite{Shibata:2013pra}), which in our notation reads~\footnote{In
  principle, there is still a weak dependence on $\alpha_0$ in ${\cal
  A}_{\beta_0}^{(A)}$.  However, this dependence becomes very small for
  $|\alpha_0| \lesssim 10^{-2}$, as it scales with $\alpha_0^2$. Therefore the
$\alpha_0$-dependence is absolutely negligible for the parameter space explored
here.}
\begin{equation}
  \alpha_A \simeq {\cal A}_{\beta_0}^{(A)}(m_A,\beta_0;{\rm EOS}) \, \alpha_0
  \;.
\end{equation}
With this equation at hand one can directly convert the limits from
ground-based GW detectors of Fig.~\ref{fig:dipole:inspiral}, for any given
${\beta_0 \gtrsim -4}$, into limits for $|\alpha_0|$ via
\begin{equation}
  |\alpha_0| = \left|\frac{\Delta\alpha}{{\cal A}_{\beta_0}^{(A)} - {\cal
  A}_{\beta_0}^{(B)}}\right| \;.
\end{equation}
Figure~\ref{fig:b0gem4} gives the results for two different mass configurations
and the EOS {\sf AP4}.  A more stiff EOS would generally lead to less
constraining limits for ground-based GW detectors and binary pulsars. As one
can see, in the range $\beta_0 \gtrsim -4$ current Solar system and pulsar
tests are already clearly more constraining than what aLIGO is expected to
obtain. For CE and ET, only inspirals with a very massive component will
provide constraints that are better than present limits, for a limited range of
$\beta_0$ (see also aLIGO~\cite{Arun:2013bp} and ET~\cite{Arun:2013bp,
Zhang:2017sym} limits from a NS-BH inspiral
for the special case of $\beta_0 = 0$, i.e.\ the Jordan-Fierz-Brans-Dicke
gravity).  By the time CE or ET is operational,
however, the expected limits from GAIA \cite{2010IAUS..261..306M} and
SKA~\cite{Berti:2015itd} will have left little room for ground-based GW
observatories in the regime. The space-based GW observatories
  LISA~\cite{Berti:2004bd, Berti:2005qd} and
  DECIGO/BBO~\cite{Yagi:2009zz} could in
principle also provide limits on the DEF theory from an inspiral of a NS into
an intermediate mass black hole, provided such BHs exist. However, the
resulting limits on $|\alpha_0|$ are not expected to be better than limits from
future ground-based GW observatories~\cite{Berti:2004bd}. It is worthy to
mention that for very large (positive) $\beta_0$, say, $\beta_0 \gtrsim
10^2\mbox{--}10^3$, massive NSs might develop
instabilities~\cite{Mendes:2014ufa, Palenzuela:2015ima}, which is beyond the
scope of Fig.~\ref{fig:b0gem4}.

%%

%-------------------------------------------------------------------------------
\subsection{Dynamical scalarization}
\label{subsec:DS}
%-------------------------------------------------------------------------------

%-------------------------------------------------------------------------------
\begin{table*}
\caption{Frequency $f_\text{DS}$ at which dynamical scalarization occurs for
various equal-mass binaries, given in Hz. Results are given for theories that
saturate the constraints given in Table~\ref{tab:psr:limit} at 68\% and 90\%
CLs.  Binary systems are specified by their component NS masses, given in units
of $M_\odot$. We highlight systems that scalarize at frequencies below
50\,Hz with boldface.}
\label{tab:DSfrequenciesEqualMass}
  \centering
  \def\arraystretch{1.2}
  \setlength{\tabcolsep}{0.42cm}
  \begin{tabularx}{\textwidth}{l@{\hskip 1.3cm}cccc@{\hskip 1.8cm}cccc}
    \hline\hline
    & \multicolumn{4}{c}{68\% confidence level} & \multicolumn{4}{c}{90\%
    confidence level} \\
    \hline
    EOS & $1.3  \enDash1.3 $ &$1.5  \enDash1.5 $  &$1.7  \enDash1.7 $  &$1.9
    \enDash1.9 $  & $1.3  \enDash1.3 $ &$1.5  \enDash1.5 $  &$1.7  \enDash1.7 $
    &$1.9  \enDash1.9 $   \\
    \hline
    {\sf AP3} & 838 & 354 & 123 & 84 & 694  & 246 & \textbf{50} & \textbf{20} \\
    {\sf AP4} & 577 & 183 & 57 & 199 & 461 & 105 &  \textbf{8}  & 109 \\
    {\sf ENG} & 858 & 358 & 118 & 102 & 694 & 236 & \textbf{39} & \textbf{24}\\
    {\sf H4} & 1301 & 650 & 235 & 51 & 1131 & 513 & 131 & \textbf{\textless 1} \\
    {\sf MPA1} & 955 & 436 & 162 & 67 & 809 & 325 & 84 & \textbf{11} \\
    {\sf MS0} & 1503 & 854 & 422 & 165 & 1320  & 700 & 302 & 81 \\
    {\sf MS2} & 1471 & 843 & 426 & 177 & 1290 & 687 & 306 & 88 \\
    {\sf PAL1} & 1350 & 693 & 287 & 95 & 1190 & 570 & 193 & \textbf{32} \\
    {\sf SLy4} & 674 & 217 & 66 & 356 & 508 & 106 & \textbf{\textless 1} & 197 \\
    {\sf WFF1} & 386 & 118 & 128 & 841 & 251 & \textbf{33} & \textbf{35} & 608 \\
    {\sf WFF2} & 519 & 154 & 57 & 302 & 391 & 72 & \textbf{\textless 1} & 181 \\
    \hline
  \end{tabularx}
\end{table*}
%-------------------------------------------------------------------------------

In addition to the nonlinear gravitational self-interaction testable with
binary pulsars, GW detectors probe the nonlinear interactions between
coalescing NSs.  Dynamical scalarization stems from the interplay between these
two regimes of strong gravity and thus offers a promising means of
complementing pulsar timing constraints on scalar-tensor theories.

Numerical relativity simulations have demonstrated that dynamical scalarization
can significantly alter the late-time behavior of a BNS system. If this
transition occurs before merger, the sudden growth of effective scalar
couplings impacts the system's gravitational binding energy and energy flux so
as to shorten the time to merger~\cite{Barausse:2012da, Shibata:2013pra,
Taniguchi:2014fqa}.

The prospective detectability of this effect was investigated in
Refs.~\cite{Sampson:2013jpa, Sampson:2014qqa} using Bayesian model selection.
The authors sought to recover injected inspiral waveforms containing dynamical
scalarization with template banks constructed from similar waveforms. The
injected signals and template banks used PN waveforms augmented with various
non-analytic models of dynamical scalarization. To mimic the abrupt activation
of the dipole emission at the onset of dynamical scalarization,
Ref.~\cite{Sampson:2013jpa} added a $-1$\,PN correction modulated by a
Heaviside function to a GR waveform, {\it i.e.}, signals of the form
\begin{align}
\tilde{h}(f)=\tilde{h}_\text{GR}(f)e^{i \Psi_\text{$-1$\,PN}(f)\Theta(f-f_*)}
\,,
\label{eq:HeavisideSignal}
\end{align}
where $\Psi_\text{$-1$\,PN}(f)=b f^{-7/3}$, and $b$ and $f_*$ are parameters of
the model. Injected signals were recovered with a template bank of waveforms of
the same form. In Ref.~\cite{Sampson:2014qqa}, the authors injected waveforms
constructed in Ref.~\cite{Palenzuela:2013hsa} by integrating the 2.5\,PN
equations of motion combined with a semi-analytic model of scalarization, then
performed parameter estimation using both templates that included $-1$\,PN and
0\,PN scalar-tensor effects throughout the entire inspiral and those that
modeled their sudden activation as in Ref.~\cite{Sampson:2013jpa}

Combined, these analyses provide a loose criterion for whether a dynamically
scalarizing BNS system could be distinguished from the corresponding system in
GR by aLIGO. The key characteristic of such systems is the frequency
$f_\text{DS}$ at which dynamical scalarization occurs. To be distinguishable
from a GR waveform, a significant portion of the dynamically scalarized
signal's SNR must occur after $f_\text{DS}$, or equivalently, $f_\text{DS}$
must be sufficiently lower than the merger frequency. Using waveforms of the
form of Eq.~(\ref{eq:HeavisideSignal}), the authors found in
Ref.~\cite{Sampson:2013jpa} that dynamical scalarization can only be observed
with aLIGO if ${f_\text{DS}\lesssim 50\enDash100}$~Hz. In only one injection
considered in Ref.~\cite{Sampson:2014qqa} was dynamical scalarization
detectable, occurring at ${f_\text{DS}\approx 80}$~Hz. Understandably, these
analyses rely on some initial assumptions that may bias these estimates away
from the real detectability criteria, such as the limited range of masses and
EOS considered and ignoring any degeneracies introduced by the merger and
ringdown portions of the waveform or by the inclusion of spins. Ignoring these
subtleties for the moment, we investigate whether the pulsar-timing constraints
described in Sec.~\ref{sec:bp} can exclude the possibility of observing
dynamical scalarization with aLIGO using the conservative detectability
criterion from Refs.~\cite{Sampson:2013jpa,Sampson:2014qqa} that scalarization
must occur by ${f_\text{DS}\lesssim50}$~Hz.

We consider binary systems composed of NSs with masses ranging from $1.3\,
M_\odot$ to $1.9\, M_\odot$. We compute $f_\text{DS}$ within the
``post-Dickean'' (PD) framework, a resummation of the PN expansion formulated
in Ref.~\cite{Sennett:2016rwa}. This model introduces new dynamical degrees of
freedom that capture the nonperturbative growth of the scalar field using a
semi-analytic feedback loop. This approach provides a mathematically consistent
backing to previous models of dynamical
scalarization~\cite{Palenzuela:2013hsa}.  The model incorporates a certain
flexibility in the choice of resummed quantities; we adopt the
$\left(m^{(\text{RE})},F^{(\tilde{\varphi})}\right)$ scheme outlined in Table I
of Ref.~\cite{Sennett:2016rwa} because it was found to give the best agreement
with numerical computations of quasi-equilibrium
configurations~\cite{Taniguchi:2014fqa}. For clarity, we dress quantities
defined in the PD framework with tildes and leave quantities defined in the PN
framework unadorned; in the limit that no resummation is performed, the PD
quantities reduce to their PN analogs.

\begin{figure*}
  \includegraphics[width=16cm]{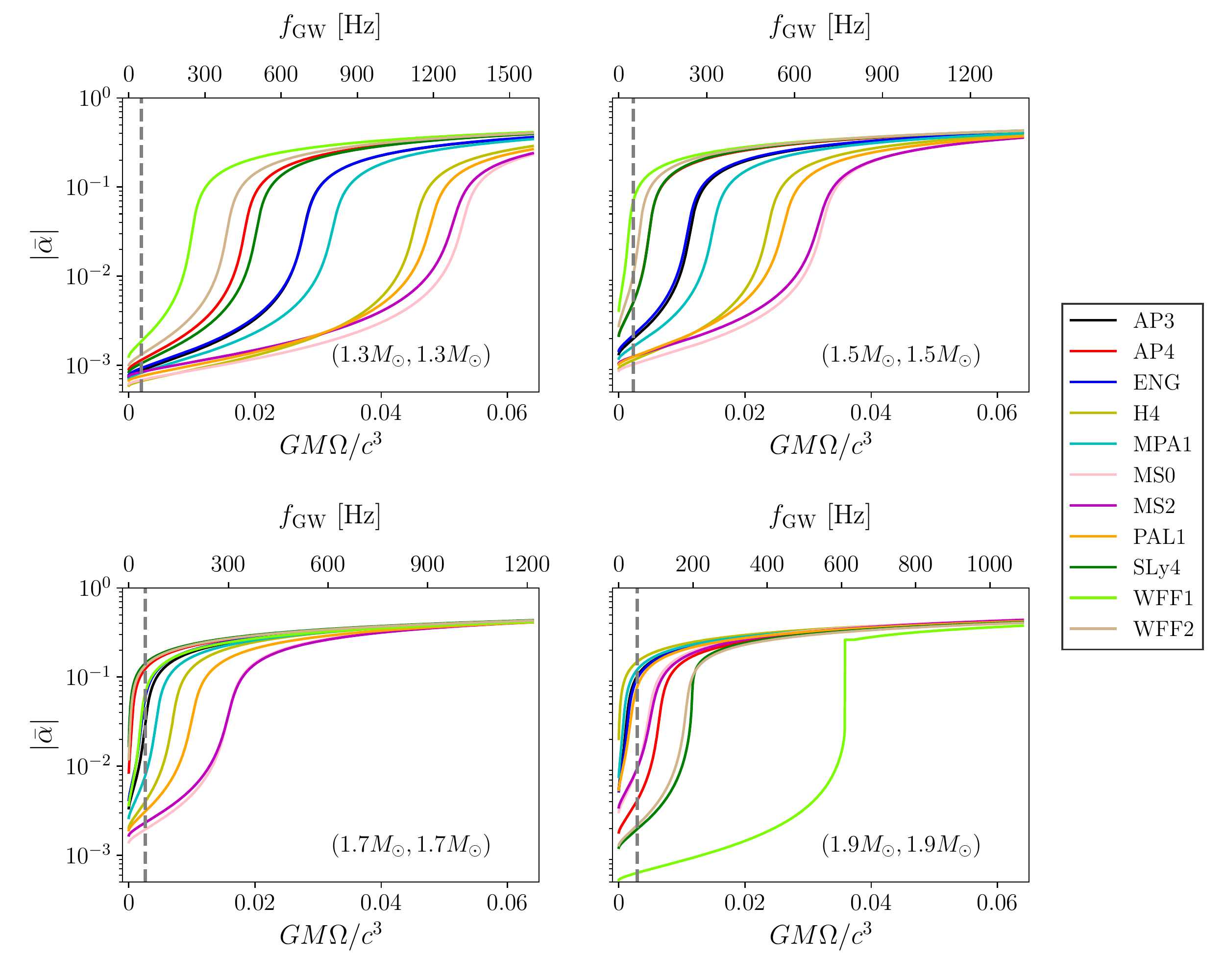}
\caption{Mass-averaged scalar coupling as a function of orbital angular
  frequency for equal-mass BNS systems with masses $\left(1.3\,M_\odot,
  1.3\,M_\odot\right)$, $\left(1.5\,M_\odot, 1.5\,M_\odot\right)$,
  $\left(1.7\,M_\odot, 1.7\,M_\odot\right)$, and $\left(1.9\,M_\odot,
  1.9\,M_\odot\right)$.  We use the limits on $(\alpha_0,\beta_0)$ at 90\% CLs,
  given in Table~\ref{tab:psr:limit}, for each EOS. The corresponding GW
  frequency is given along the top axis, with $f_\text{GW}=\Omega/\pi$. Dashed
  vertical lines highlight the conservative detectability criterion for aLIGO
  that $f_\text{DS}\lesssim50\,{\rm Hz}$, derived in
  Refs.~\cite{Sampson:2013jpa, Sampson:2014qqa}.
  \label{fig:DS}}
\end{figure*}
%%

%-------------------------------------------------------------------------------
\begin{table*}
\caption{Frequency $f_\text{DS}$ (in Hz) at which dynamical scalarization
occurs for various unequal-mass binaries with the EOS {\sf MPA1}. Results are
given for theories that saturate the constraints given in
Table~\ref{tab:psr:limit} at 68\% and 90\% CLs.  Binary systems are specified
by their component NS masses, given in units of $M_\odot$. We highlight systems
that scalarize at frequencies below 50 Hz with boldface.}
\label{tab:DSfrequenciesUnequalMass}
  \centering
  \def\arraystretch{1.2}
  \setlength{\tabcolsep}{0.22cm}
  \begin{tabularx}{\textwidth}{c@{\hskip 1cm}cccccccc@{\hskip 1cm}cccccccc}
    \hline\hline
        & \multicolumn{8}{c}{68\% confidence level} & \multicolumn{8}{c}{90\%
    confidence level} \\
	\hline
       & 1.2 & 1.3 & 1.4 & 1.5 & 1.6 & 1.7 & 1.8 & 1.9 & 1.2 & 1.3 & 1.4 & 1.5 &
       1.6 & 1.7 & 1.8 & 1.9 \\
       \hline
    1.2 & 1340 & 1130 & 942 & 767 & 611 & 471 & 364 & 304 & 1160 & 973 & 788 &
    618 & 458 & 315 & 194 & 119 \\
    1.3 & \ldots & 955 & 795 & 649 & 515 & 399 & 306 & 258 & \ldots & 809 & 653
    & 511 & 380 & 262 & 158 & 99 \\
    1.4 & \ldots & \ldots & 661 & 538 & 427 & 329 & 253 & 212 & \ldots & \ldots
    & 529 & 415 & 305 & 210 & 129 &  79 \\
    1.5 & \ldots & \ldots & \ldots & 436 & 347 & 268 &  206 & 172 & \ldots &
    \ldots & \ldots & 325 & 240 & 165 & 101 & 62 \\
    1.6 & \ldots & \ldots & \ldots & \ldots & 274 & 212 & 163 & 136 & \ldots &
    \ldots & \ldots & \ldots & 178 & 120 & 73 & \textbf{46} \\
    1.7 & \ldots & \ldots & \ldots & \ldots & \ldots & 162 & 126 & 105 & \ldots
    & \ldots & \ldots & \ldots & \ldots & 84 & 51 & \textbf{31} \\
    1.8 & \ldots & \ldots & \ldots & \ldots & \ldots & \ldots & 96 & 80 & \ldots
    & \ldots & \ldots & \ldots & \ldots & \ldots & \textbf{32} & \textbf{19} \\
    1.9 & \ldots & \ldots & \ldots & \ldots & \ldots & \ldots & \ldots & 67 &
    \ldots & \ldots & \ldots & \ldots & \ldots & \ldots & \ldots & \textbf{11} \\
    \hline
  \end{tabularx}
\end{table*}
%-------------------------------------------------------------------------------

Within the PD framework, the effective scalar coupling of each NS is
promoted to a function of both the asymptotic scalar field $\varphi_0$ and the
local scalar field in which the body is immersed, i.e.  $\tilde\alpha_A=\tilde
\alpha_A(\varphi_0,\varphi_A)$. Unlike in the PN treatment, this coupling
evolves as the BNS coalesces. Similarly, the inertial mass of each body $\tilde
m_A(\varphi_0,\varphi_A)$ evolves in the PD framework. However this mass varies
by no more than $0.01\%$, so in practice, one can simply use the PN mass $m_A$
in place of $\tilde m_A$.

We define the mass-averaged scalar coupling of the system as
\begin{equation}
\bar\alpha\equiv\frac{\tilde m_A \tilde \alpha_A+\tilde m_B \tilde
\alpha_B}{\tilde m_A+\tilde m_B},
\end{equation}
where precise definitions of $\tilde m_A$ and $\tilde \alpha_A$ are given in
Eqs.~(A3) and~(A4) in Ref.~\cite{Sennett:2016rwa}. Note that for equal-mass
binaries, we have ${\tilde\alpha_A=\tilde\alpha_B=\bar \alpha}$.

Following the work of Ref.~\cite{Sennett:2016rwa}, we compute the
mass-averaged scalar coupling as a function of frequency for binaries on
quasi-circular orbits to 1\,PD order. The average scalar coupling is plotted
in Fig.~\ref{fig:DS} for equal-mass binaries for theories that saturate the
pulsar-timing constraints at 90\% CLs, as given in Table \ref{tab:psr:limit}.
Scalarization occurs earlier for larger mass systems, with an ordering (by EOS)
determined by the magnitude of
\begin{equation}
  \beta_A=\left(\frac{{\rm d} \alpha_A}{ {\rm d}
\varphi}\right)_{\varphi=\varphi_0}\,.
\label{eq:betaAdef}
\end{equation}
To compute this quantity, one takes the difference in effective scalar
couplings of NSs (of equal baryonic mass) with infinitesimally different
asymptotic scalar fields $\varphi_0$; however, for non-spontaneously scalarized
stars, $\beta_A$ is given approximately by
\begin{equation}
\beta_A\approx \frac{\beta_0 |\alpha_A|}{|\alpha_0|} \,,
\end{equation}
provided that $|\alpha_A|$ is sufficiently small. Binaries with spontaneously
scalarized stars begin with an appreciable effective scalar coupling at
large separations that continues to grow as they coalesce. In light of this
remark, we note that there is no observational distinction between spontaneous
(or induced) scalarization and dynamical scalarization that occurs at
sufficiently low frequencies; for example, compare the scalarization of
$\left(1.7\,M_\odot, 1.7\,M_\odot\right)$ systems composed of NSs with the EOSs
{\sf SLy4}, {\sf AP4}, and {\sf WFF1} (the dark green, red, and beige curves in
the lower left panel of Fig.~\ref{fig:DS}, respectively).

The sharp feature for the {\sf WFF1} EOS in the $\left(1.9\,M_\odot,
1.9\,M_\odot\right)$ system occurs because of the relatively low mass at which
spontaneous scalarization occurs for this particular EOS. We provide a more
detailed analysis of this phenomenon in Appendix~\ref{sec:MassiveBNSDS}.
Similarly abrupt transitions occur for other EOSs in more massive binary
systems with individual masses $ \gtrsim 2\,M_\odot$.

We adopt the method introduced in Ref.~\cite{Taniguchi:2014fqa} to extract
$f_\text{DS}$. The average effective scalar coupling can be closely fit by
the piecewise function
\begin{equation}
  \label{eq:fitEq}
\left(1+\bar\alpha^2\right)^{10/3}=1+a_1\left(x-x_\text{DS}\right)
\Theta\left(x-x_\text{DS}\right)
\end{equation}
where $\Theta$ is the Heaviside function, $a_1$ and $x_\text{DS}$ are fitting
parameters, and $x\equiv(G_* M \Omega/c^3)^{2/3}$. In practice, we identify
$x_\text{DS}$ with the peak in the second derivative of the lefthand side of
Eq.~(\ref{eq:fitEq}) with respect to $x$. The gravitational wave frequency at
which dynamical scalarization occurs is then given by
$f_\text{DS}=\Omega_\text{DS}/\pi$. In Ref.~\cite{Sennett:2016rwa}, the PD
prediction was found to reproduce numerical-relativity results to within an
error of $\lesssim 10 \%$ with this fitting procedure.

The dynamical scalarization frequencies for the configurations considered in
Fig.~\ref{fig:DS} are given in Table~\ref{tab:DSfrequenciesEqualMass} for
theories constrained at the 68\% and 90\% CLs.  Systems containing
spontaneously scalarized stars ({\it i.e.}, those with appreciable effective
scalar coupling even in isolation) are demarcated as scalarizing below 1\,Hz;
as noted above, these systems would be indistinguishable to GW detectors from
those that dynamically scalarize below 1\,Hz. For clarity, we highlight the
systems in Table~\ref{tab:DSfrequenciesEqualMass} that scalarize (dynamically
or spontaneously) below 50 Hz. Recall that, under our definition, induced
scalarization occurs in binaries comprised of one initially scalarized star and
one initially unscalarized star; this asymmetry cannot be achieved in
equal-mass systems like those discussed above.

We next consider the onset of dynamical scalarization in unequal-mass systems.
For the sake of compactness, we show in
Table~\ref{tab:DSfrequenciesUnequalMass} the dynamical scalarization
frequencies for binaries with NS masses of $1.2\, M_\odot$ to $1.9\, M_\odot$
with just the {\sf MPA1} EOS. We find that the total mass plays a more
important role in determining the onset of dynamical scalarization than the
mass ratio. Fixing the total mass, we find that scalarization occurs earlier in
more asymmetric binaries of lower mass ({\it e.g.}, ${M \lesssim 3.2M_\odot}$
for the {\sf MPA1} EOS). None of the systems listed in
Table~\ref{tab:DSfrequenciesUnequalMass} undergo induced scalarization. As
before, we highlight the systems in Table~\ref{tab:DSfrequenciesUnequalMass}
that scalarize below 50 Hz.

To summarize, Tables~\ref{tab:DSfrequenciesEqualMass}
and~\ref{tab:DSfrequenciesUnequalMass} demonstrate that binary-pulsar
constraints cannot entirely rule out the possibility of dynamical scalarization
occurring at frequencies $f_\text{DS}\lesssim 50$~Hz at 90\% CL. Initial
detectability studies --- Refs.~\cite{Sampson:2013jpa,Sampson:2014qqa}
discussed above --- suggest that this early scalarization should be observable
with aLIGO (although these conclusions should be confirmed with future work in
light of the limitations of these works; see above). Thus, failure to detect
dynamical scalarization in future GW observations could provide tighter
constraints on the parameters $(\alpha_0, \beta_0)$ in DEF theory than pulsar
timing. However, as can be seen from Table~\ref{tab:DSfrequenciesEqualMass},
the prospects of producing such complementary constraints depend critically on
the observed NS masses and the EOS of NSs.

%===============================================================================
\section{Conclusions}
\label{sec:con}
%-------------------------------------------------------------------------------

In this paper, we have studied the scalarization phenomena~\cite{Damour:1993hw,
Barausse:2012da} in the massless mono-scalar-tensor theory of gravity of DEF
with pulsar timing and laser-interferometer GW detectors on the Earth. We now
summarize the key conclusions of our analysis.
%-------------------------------------------------------------------------------
\begin{enumerate}
  \item The spontaneous scalarization phenomenon~\cite{Damour:1993hw} occurs at
    different NS mass ranges for different EOSs~\cite{Shibata:2013pra}.
    Therefore, in a well-timed relativistic binary-pulsar system with a
    specific NS mass, the scalar-tensor gravity might be stringently
    constrained for some EOSs whose spontaneous-scalarization phenomenon occurs
    near that specific NS mass. However, in general, strong scalarization could
    still take place if NSs are described by an EOS whose scalarization occurs
    at a mass different from the one observed.
  \item Combining two well-timed binary-pulsar systems with quite different NS
    masses, one could in principle constrain the scalar-tensor theory with
    whatever EOS Nature provides us. Using MCMC simulations, we showed in
    Sec.~\ref{sec:bp} that, combining five binary
    pulsars~\cite{Lazaridis:2009kq, Freire:2012mg, Antoniadis:2013pzd,
    Reardon:2015kba, Cognard:2017xyr} that best constrain gravitational dipolar
    radiation, we can already bound the {\it scalarization parameter},
    $\beta_0$, to be $\gtrsim -4.28$ at 68\% CL and $\gtrsim -4.38$ at 90\% CL,
    for any of the eleven EOSs that we have considered.
  \item Nevertheless, because of the limited distribution of masses of the five
    chosen binary pulsars, we found that if the EOS of NSs were similar to one
    of {\sf AP4}, {\sf SLy4}, or {\sf WFF2}, NSs with masses of $m_A \simeq
    1.70$--$1.73\,M_\odot$ could still develop an effective scalar coupling
    $\gtrsim {\cal O}(10^{-2})$.  This is also true for the EOS {\sf H4} with
    $m_A \simeq 1.92\,M_\odot$, and for the EOSs {\sf MS0} and {\sf MS2} with
    $m_A \simeq 2.26\,M_\odot$ (see Fig.~\ref{fig:max:alphaA}).
  \item Using the upper limits on the effective scalar coupling of NSs from
    binary pulsars, we found that for BNSs in  the frequency bands of aLIGO,
    CE, and ET,  we could still have a large time-domain dephasing in the
    number of GW cycles, on the orders of ${\cal O}(10^1)$, ${\cal O}(10^2)$,
    and ${\cal O}(10^3)$, respectively (see Table~\ref{tab:DelN}).
  \item We performed a Fisher-matrix study of BNS inspiral signals using aLIGO,
    CE, and ET. We found that for BNSs at a luminosity distance $D_L=200$\,Mpc,
    where we expect to observe those sources, aLIGO can still improve the
    limits from binary pulsars for a couple of EOSs with BNSs of suitable
    masses. CE (whose bandwith starts at 5\,Hz) can improve the current limits
    for all EOSs, while ET (whose bandwith starts at 1\,Hz) will provide us
    with even more significant improvements over current constraints for all
    EOSs. This is mainly due its better low-frequency sensitivity.  Our
    conclusions for aLIGO differ from the one obtained in
    Refs.~\cite{Damour:1998jk, EspositoFarese:2004tu, EspositoFarese:2004cc},
    where the authors concluded that pulsar timing would do better than aLIGO
    in constraining scalar-tensor theories.  The main reason of this difference
    comes from a better understanding and larger span of the NS masses and EOSs
    during the past two decades~\cite{Demorest:2010bx, Antoniadis:2013pzd}, and
    the different PSD for aLIGO used in Refs.~\cite{Damour:1998jk,Will:1994fb}.
    If we restricted the analysis to the same NS masses and the same EOS used
    in Ref.~\cite{Damour:1998jk}, we would recover the same conclusions as in
    Ref.~\cite{Damour:1998jk} (see Fig.~\ref{fig:dipole:inspiral}).
  \item We investigated dynamical scalarization in equal-mass and unequal-mass
    BNS systems. With the criterion that the dynamical scalarization transition
    frequency must fall below $\sim50$\,Hz~\cite{Sampson:2013jpa,
    Sampson:2014qqa} to be detectable, we found aLIGO could be able to observe
    this phenomenon given the constraints obtained from binary-pulsar timing,
    even away from the scalarization windows.  We found that the prospects
    for observing dynamical scalarization with GW detectors depends critically
    on the NS EOS---for example, dynamical scalarization of NSs with the {\sf
    MS0} EOS could not be detected with aLIGO. Producing new constraints on
    scalar-tensor theories from GW searches for dynamical scalarization
    requires waveform models that can faithfully reproduce this nonperturbative
    phenomenon; ultimately, these conclusions should be revisited once such
    models are developed.

\end{enumerate}
%-------------------------------------------------------------------------------

Our comparisons between binary pulsars and GWs made use of the {\it current}
limits of the former and the {\it expected} limits of the latter. It shows that
advanced and next-generation ground-based GW detectors have potential to
further improve the current limits set by pulsar timing. Nevertheless, the
binary-pulsar limits will also improve over time, especially if suitable
systems filling the scalarization windows are discovered in future pulsar
surveys. Better mass measurements of currently known pulsars will also help in
narrowing down the constraints, especially with
PSRs~J1012+5307~\cite{Lazaridis:2009kq} and J1913+1102~\cite{Lazarus:2016hfu},
whose observational uncertainties in masses are still large, and they might
have the right masses to close the windows below $2\,M_\odot$. To reach
this goal, the next generation of radio telescopes, such as FAST and SKA will
play a particularly important role \cite{Nan:2011um,Shao:2014wja}.

%===============================================================================

\begin{acknowledgments}
We thank Paulo Freire, Ian Harry, Jan Steinhoff, Thomas Tauris, and Nicol\'as
Yunes for helpful discussions.  We are grateful to Jim Lattimer for providing
us with tabulated data for neutron-star equations of state.  The Markov-chain
Monte Carlo runs were performed on the {\sc Vulcan} cluster at the Max Planck
Institute for Gravitational Physics in Potsdam.
\end{acknowledgments}

%% === END of main body of the paper ===

\appendix

%-------------------------------------------------------------------------------
\section{Ingredients for the Fisher matrix analysis
\label{sec:app:fisher}}
%-------------------------------------------------------------------------------

For a GW detector with the one-side PSD, $S_n(f)$, the SNR of a Fourier-domain
waveform, $\tilde h(f)$, is
\begin{equation}\label{eq:snr}
  \rho = \left( \tilde h(f) \right| \left. \tilde h(f) \right)^{1/2} \,,
\end{equation}
where the inner product is defined to be~\cite{Finn:1992wt, Cutler:1994ys},
\begin{equation}
  \label{eq:inner:prod}
  \left(\tilde h_1(f) \right| \left. \tilde h_2(f) \right) \equiv 2
  \int_{f_{\rm min}}^{f_{\rm max}} \frac{\tilde h_1^*(f) \tilde h_2(f) + \tilde
  h_1(f) \tilde h^*_2(f)}{S_n(f)} {\rm d} f \,.
\end{equation}
For all calculations in Sec.~\ref{subsec:fisher}, we use the design
zero-detuned high-power noise PSD, starting from $10\,$Hz for
aLIGO~\cite{Shoemaker:2010}, the target noise PSD, starting from $5\,$Hz for
CE~\cite{Evans:2016mbw}, and the {\sc ET-D} noise PSD, starting from $1\,$Hz
for ET~\cite{Hild:2010id}.  Thus, in Eq.~(\ref{eq:inner:prod}) we choose
$f_{\rm min}=10\,$Hz, $5\,$Hz, and $1\,$Hz for aLIGO, CE, and ET, respectively.
Somewhat arbitrarily, we choose for $f_{\rm max}$ twice the innermost stable
circle orbit (ISCO) frequency computed from the binary's binding energy at
2\,PN order. It reads~\cite{Favata:2010yd},
\begin{equation}
  \label{eq:isco}
  f_{\rm max} = \frac{c^3}{\pi G M} \left[ \frac{3}{14\eta} \left(
  1-\sqrt{1-\frac{14}{9}\eta} \right) \right]^{3/2} \,.
\end{equation}
Because the three detectors do not have good sensitivity at high frequency, say
$\gtrsim$\,kHz, the choice of $f_{\rm max}$ influences the result very
marginally.

For the nonspinning BNS inspiraling waveform in the Fourier domain, we use a
restricted waveform with the leading-order term in amplitude, ${\cal A}$, and
up to 3.5\,PN terms in the phase, $\Psi(f)$~\cite{Will:1994fb, Berti:2004bd,
Buonanno:2009zt},
\begin{widetext}
\begin{eqnarray}
  \tilde h(f) &=& {\cal A} f^{-7/6} e^{i \Psi(f)} \,,
  \label{eq:fd:wf} \\
  \Psi(f) &=&  2\pi f t_c - \Phi_c - \frac{\pi}{4} + \frac{3}{128\eta}
  \mathfrak{u}^{-5/3} \left\{ 1 - \frac{5}{168} (\Delta\alpha)^2
  \mathfrak{u}^{-2/3} + \left( \frac{3715}{756} + \frac{55}{9} \eta \right)
  \mathfrak{u}^{2/3} - 16\pi \mathfrak{u}  \right. \nonumber \\
    && \left. +  \left( \frac{15293365}{508032} + \frac{27145}{504}\eta +
    \frac{3085}{72}\eta^2 \right) \mathfrak{u}^{4/3} + \pi \left(
    \frac{38645}{756} - \frac{65}{9} \eta \right) \left( 1 + \ln
    \mathfrak{u} \right) \mathfrak{u}^{5/3} \right. \nonumber\\
    && \left. + \left[ \frac{11583231236531}{4694215680} - \frac{640}{3}\pi^2 -
    \frac{6848}{21} \gamma_{\rm E} -\frac{6848}{63}\ln\left( 64\mathfrak{u}
    \right) + \left( -\frac{15737765635}{3048192} + \frac{2255}{12}\pi^2
    \right)\eta + \frac{76055}{1728}\eta^2 \right.\right. \nonumber \\
    && \left.\left. - \frac{127825}{1296}\eta^3 \right] \mathfrak{u}^2 + \pi
    \left( \frac{77096675}{254016} + \frac{378515}{1512}\eta -
    \frac{74045}{756}\eta^2 \right) \mathfrak{u}^{7/3} \right\} \,,
    \label{eq:fd:wf:phase}
\end{eqnarray}
\end{widetext}
where $\mathfrak{u} \equiv \pi G M f / c^3$, ${\cal A} \propto {\cal M}^{5/6} /
D_L$ with the chirp mass ${\cal M}\equiv\eta^{3/5} M$ and the luminosity
distance $D_L$, $t_c$ and $\Phi_c$ are reference time and phase
respectively, and $\gamma_{\rm E} = 0.577216\dots$ is the Euler constant.
Note that in Eq.~(\ref{eq:fd:wf:phase}) the gothic $\mathfrak{u}$
is equal to $\eta^{-3/5}u$ where  $u\equiv \pi G {\cal M}f / c^3$, as defined
in Ref.~\cite{Will:1994fb}.  We include in Eq.~(\ref{eq:fd:wf:phase}) only the
leading dipole term for the scalar contribution.  Furthermore, since the spins
of BNS systems are supposed to be small, we do not include them in the analysis
(see Table~\ref{tab:DelN} where we give a rough estimation of the spin terms in
the GW phasing).

To calculate the Fisher matrix (\ref{eq:fisher:matrix}), we need to compute
partial derivatives of the frequency-domain waveform~(\ref{eq:fd:wf}). They
read (notice that, when calculating derivatives, $\mathfrak{u}$ depends on both
$\eta$ and ${\cal M}$),
\begin{widetext}
\begin{eqnarray}
  \frac{\partial \tilde h(f)}{\partial \ln {\cal A}} &=& \tilde h(f) \,,\\
  \frac{\partial \tilde h(f)}{\partial \ln \eta} &=& \frac{i}{\eta}
  \mathfrak{u}^{-5/3} \left\{ -\frac{1}{3584 } \Delta \alpha ^2
  \mathfrak{u}^{-2/3} + \left( -\frac{743}{16128}+\frac{11 }{128} \eta \right)
  \mathfrak{u}^{2/3} + \frac{9 }{40 } \pi  \mathfrak{u} +
  \left(-\frac{3058673}{5419008}  +\frac{5429}{21504}\eta  + \frac{617}{512}
  \eta^2 \right) \mathfrak{u}^{4/3}  \right.  \\
    && \left. + \pi \left( -\frac{7729}{4032} -\frac{38645 }{32256}\ln
    \mathfrak{u} + \frac{13 }{128} \eta \right) \mathfrak{u}^{5/3} + \left[
      -\frac{11328104339891}{166905446400}+6 \pi ^2  + \frac{321
      }{35}\gamma_{\rm E} + \frac{107}{35} \ln\left( 64 \mathfrak{u}
      \right)\right.  \right. \nonumber \\
    && \left. \left.  + \left( \frac{3147553127}{130056192}-\frac{451
    }{512}\pi^2\right) \eta  + \frac{15211}{18432} \eta ^2 -\frac{25565}{6144}
  \eta ^3 \right] \mathfrak{u}^{2} + \pi
  \left(-\frac{15419335}{1548288}-\frac{75703 }{32256} \eta -\frac{14809
  }{10752} \eta ^2\right) \mathfrak{u}^{7/3} \right\} \tilde h(f)\nonumber \,,\\
  \frac{\partial \tilde h(f)}{\partial \ln {\cal M}} &=& \frac{i}{\eta}
  \mathfrak{u}^{-5/3} \left\{
    -\frac{5}{128} +\frac{5}{3072}\Delta \alpha^2\mathfrak{u}^{-2/3}
  + \left(-\frac{3715}{32256}-\frac{55}{384}\eta\right) \mathfrak{u}^{2/3}
  +\frac{\pi}{4} \mathfrak{u} + \left(-\frac{15293365}{65028096}
  -\frac{27145}{64512}  \eta  -\frac{3085 }{9216}\eta ^2 \right)
  \mathfrak{u}^{4/3} \right. \nonumber \\
  && \left. + \pi\left(\frac{38645 }{32256}-\frac{65 }{384} \eta \right)
  \mathfrak{u}^{5/3} + \left[
    \frac{10052469856691}{600859607040}-\frac{5}{3}\pi^2
    -\frac{107}{42}\gamma_{\rm E} -\frac{107}{126}\ln\left( 64\mathfrak{u}
    \right) + \left(-\frac{15737765635}{390168576} +
    \frac{2255}{1536}\pi^2\right)\eta \right. \right. \nonumber \\
    && \left. \left. + \frac{76055}{221184} \eta^2-\frac{127825}{165888} \eta^3
  \right] \mathfrak{u}^2 + \pi \left(\frac{77096675}{16257024}
  +\frac{378515}{96768} \eta  -\frac{74045 }{48384}\eta ^2\right)
  \mathfrak{u}^{7/3} \right\}\tilde h(f) \,, \\
  \frac{\partial \tilde h(f)}{\partial t_c} &=& i2\pi f \tilde h(f) \,,\\
  \frac{\partial \tilde h(f)}{\partial \Phi_c} &=& -i \tilde h(f) \,, \\
  \frac{\partial \tilde h(f)}{\partial (\Delta\alpha)^2} &=& -
  i\frac{5}{7168\eta} \mathfrak{u}^{-7/3}  \tilde h(f)  \,.
\end{eqnarray}
\end{widetext}

%-------------------------------------------------------------------------------
\section{Dynamical scalarization in ultra-relativistic binary neutron stars}
\label{sec:MassiveBNSDS}
%-------------------------------------------------------------------------------

In this appendix, we discuss the sharp feature observed in the averaged
effective scalar coupling of a very massive BNS that undergoes dynamical
scalarization (see the $1.9 \,M_\odot \enDash 1.9\,M_\odot$ case in
Fig.~\ref{fig:DS}). We find that generically NSs of very high mass can
scalarize more abruptly than their less massive counterparts.

From Fig.~\ref{fig:max:alphaA}, we observe that very massive NSs exhibit very
small effective scalar couplings $\alpha_A$. In these stars, the effective
scalar coupling is nonperturbatively suppressed below the non-relativistic
(low-mass) limit $\alpha_A\approx \alpha_0$. The cores of these stars are
ultra-relativistic, with a negative trace of the stress-energy tensor
$T_*=\epsilon_* -3 p_* <0$.  The mass at which NSs become ultra-relativistic in
this sense depends on the EOS and can be read off from
Fig.~\ref{fig:dipole:inspiral} as the mass at which the best constraint on
$|\alpha|$ drops to zero. Recall that spontaneous scalarization stems from a
large, positive source on the righthand side of Eq.~(\ref{eq:field:scalar})
that grows with $\varphi$. In ultra-relativistic stars, this source term
becomes negative, causing the star to spontaneously ``de-scalarize''.

\begin{figure}[t]
 \includegraphics[width=9cm]{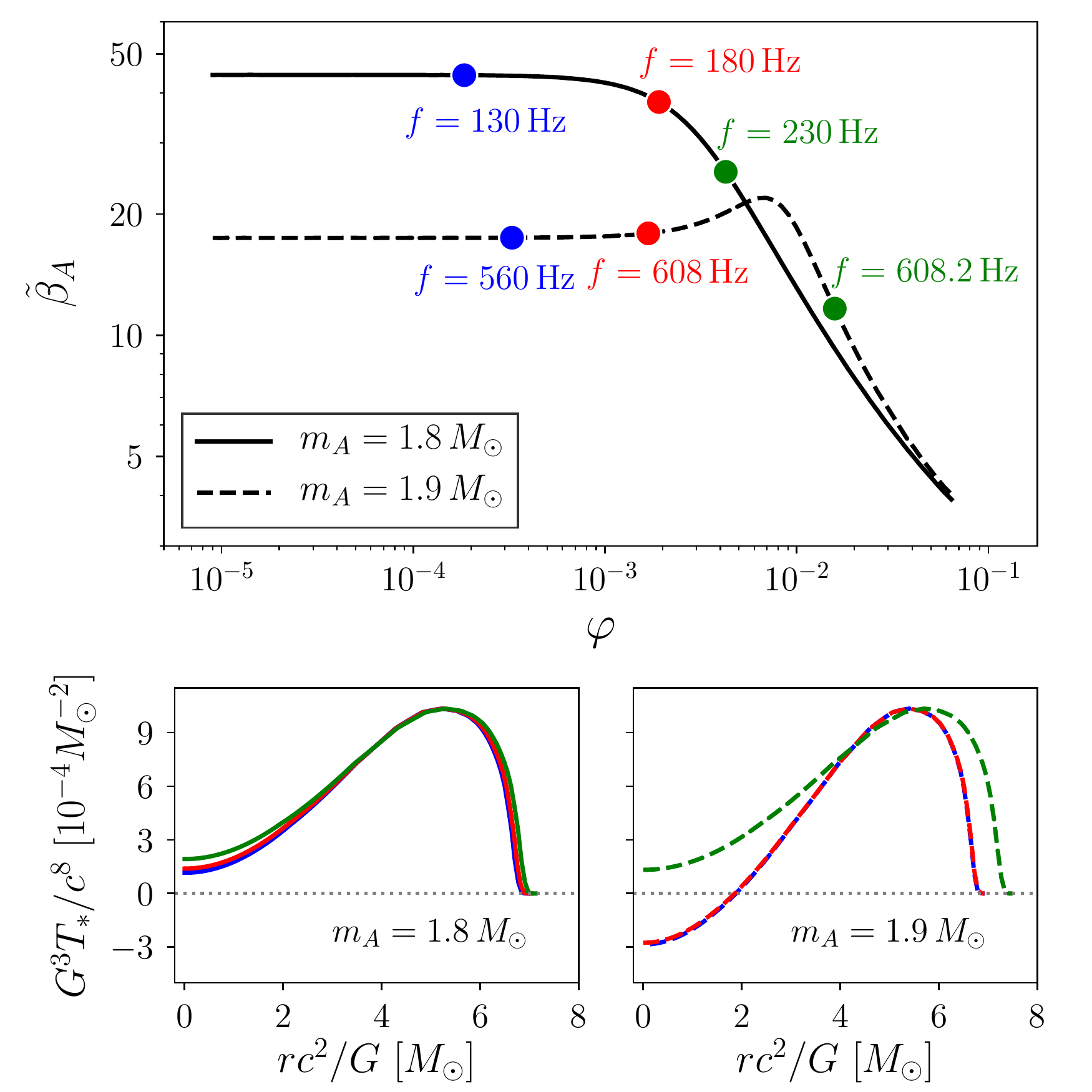}
 \caption{({\it Top}) Evolution of $\tilde\beta_A$ for NSs with (dashed) and
 without (solid) ultra-relativistic cores as a function of the background
 scalar field.  The blue, red, and green annotations indicate the scalar field
 at each star at frequencies before, during, and after dynamical scalarization
 in an equal-mass binary system respectively. ({\it Bottom}) Profile of the
 trace of the stress-energy tensor within each NS at each of the annotated
 points in the top panel.}
 \label{fig:DS:pt}
\end{figure}

When placed in a binary system, ultra-relativistic NSs can dynamically
scalarize, but the transition occurs very abruptly ({\it e.g.}, the $1.9\,
M_\odot \enDash 1.9\, M_\odot$ system with the EOS {\sf WFF1} shown in
Fig.~\ref{fig:DS}). As the system scalarizes, the massive NSs transition to a
state in which $T_*$ is everywhere positive. Figure~\ref{fig:DS:pt} depicts
this transition in comparison to dynamical scalarization in less massive
systems. The top panel shows $\tilde\beta_A$---the PD equivalent of the
quantity defined in Eq.~(\ref{eq:betaAdef})---for $1.8 \, M_\odot$ (solid) and
$1.9\, M_\odot$ (dashed) stars with the EOS {\sf WFF1} plotted as a function of
scalar field. The highlighted points indicate the field at each NS in an
equal-mass binary before (blue), during (red), and after (green) dynamical
scalarization. The bottom panels depict the profile of $T_*$ within each star
at each of these points.

The top panel of Fig.~\ref{fig:DS:pt} demonstrates why dynamical scalarization
occurs abruptly for ultra-relativistic NSs. Recall that $\beta_A$ (and
consequently $\tilde\beta_A$) quantifies how easily a NS can scalarize, with
larger values indicating that the star is more susceptible to dynamical
scalarization. As expected, when immersed in a weak scalar field ({\it i.e.},
during the early inspiral), $\tilde\beta_A$ is significantly smaller for
ultra-relativistic stars than less massive stars, indicating that the latter
will dynamically scalarize at a lower frequency.  However, unlike for less
massive stars, $\tilde\beta_A$ increases slightly as the scalar field reaches
larger values ($\varphi\sim0.002$ in Fig.~\ref{fig:DS:pt}) for
ultra-relativistic NSs. This triggers a run-away process in a binary system as
a small increase in the scalar field produced by one star causes the other star
to scalarize more easily, which in turn allows the second star to produce a
larger scalar field for the first. For the ultra-relativistic BNS depicted in
the top panel of Fig.~\ref{fig:DS:pt}, this transition is completed after an
evolution of only $0.2$ Hz.

%\bibliography{refs.bib}

%

\end{document}